\documentclass[a4paper,11pt]{article}

\usepackage{jcappub}
\usepackage[T1]{fontenc} 
\usepackage{graphicx}
\usepackage{amsmath}
\usepackage{amssymb}
\usepackage{amsfonts}
\usepackage{lscape}
\usepackage{color}
\usepackage{bm}

\def\be{\begin{equation}}
\def\ee{\end{equation}}
\def\bea{\begin{eqnarray}}
\def\eea{\end{eqnarray}}
\newcommand{\comment}[1]{}

\newcommand{\average}[1]{\left\langle #1 \right\rangle}

\title{Observational tests of backreaction with recent data}
\abstract{something something something}
\notoc

\begin{document}

\author[a,b]{Matteo Chiesa}
\author[a]{Davide Maino}
\author[c,b]{Elisabetta Majerotto}

\affiliation[a]{University of Milan, Physics Department, \\ via Giovanni Celoria 16, Milan, Italy}
\affiliation[b]{INAF-Osservatorio Astronomico di Brera, \\ via Emilio Bianchi 46, Merate, Italy}
\affiliation[c]{Departamento de F\'isica Te\'orica and Instituto de F\'isica Te\'orica, \\
Universidad Aut\'onoma de Madrid IFT-UAM/CSIC,\\
$28049$ Cantoblanco, Madrid, Spain}

\emailAdd{matteo.chiesa@unimi.it, matteo.chiesa.84@gmail.com}
\emailAdd{davide.maino@mi.infn.it}
\emailAdd{elisabetta.majerotto@uam.es}

\date{\today}

\abstract{We explore the backreaction model based on the template metric proposed in Larena et al. \cite{Larena:2008be} constraining the matter density parameter $\Omega_m^{D_0}$ and the Dark Energy parameter $w$ with recent data. We provide constraints based on Supernovae Ia from the \emph{SNLS} and the \textit{Union2.1}  catalogs, confirming that the backreacted Universe should have a higher matter density than the corresponding Friedmaniann one. Angular diameter distances from clusters data confirm the same feature. Finally we combine these results with constraints obtained from the position of the first three peaks and the first dip of the  CMB power spectrum, fitting \emph{WMAP-9}  and \emph{Planck}  data. We find that an inconsistency arises in predicting the scale factor at recombination, leading to a backreacted Universe with lower matter density, in contradiction with results produced by SnIa and clusters. The same behavior is confirmed by analyzing the CMB-shift parameters from \emph{WMAP-9}. We conclude exploring qualitatively the motivations of this inconsistency.}

\maketitle

\section{Introduction}

The Universe on large scales appears to be homogeneous and isotropic. General Relativity (GR) provides a simple and clear description of the Universe in terms of a homogeneous and isotropic spacetime metric: the Friedmann-Lemaitre-Robertson-Walker (FLRW) metric. 

The accelerated expansion we can observe from supernovae type Ia (SnIa) \cite{Riess:1998cb,Perlmutter:1998np} contrasts such a simple view of our Universe: ordinary types of matter and radiation lead to a deceleration of the expansion, and a new negative pressure fluid is needed to explain the acceleration. This fluid, called Dark Energy (DE), should account for about 70\% of the whole energy content of the Universe. (see for example \cite{Seljak:2004xh,Riess:2004nr} and references therein) 

Many different models have been built to describe the physical features of DE (see e.g \cite{Copeland:2006wr,Frieman:2008sn,Li:2011sd} for a review). The simplest one is based on a cosmological constant which enters Einstein equations, while other models assume some exotic new energy content such as a scalar field called quintessence. The apparent acceleration could also be described by a modification of GR on large scales. The debate between DE and Modified Gravity models is still open.

Recently it has been proposed to describe the cosmic acceleration as an effect of the backreaction of the local inhomogeneities (see \cite{Buchert:1995fz,Rasanen:2003fy,Rasanen:2006kp,Kolb:2004am,Kolb:2005da} and references therein).
Indeed the procedure of averaging the local inhomogeneties over a finite domain, and its global dynamics are non-commutative operations which modify the Friedmann equations, allowing in principle for an apparent accelerated expansion without the need of postulating any new cosmological fluid (see \cite{Buchert:2007ik} for an overview). 

Models of backreaction were first built in the context of Newtonian gravity \cite{Buchert:1995fz,Buchert:1999pq}. Generalizations
 followed \cite{Buchert:1999er,Buchert:2001sa,Buchert:2002ht}, focusing on the interpretation of the averaging problem in the context of General Relativity. In these models the spacetime metric plays a crucial role: the full metric which describes the non linear inhomogeneities and their gravitational effects is unknown, requiring to make strong assumptions on its form. For instance a perturbative approach could be used up to linear scales, starting from a homogeneous FLRW background, but non-perturbative schemes have been suggested, too \cite{Paranjape:2006ww}.

Larena et al. \cite{Larena:2008be} performed a test on a backreaction model based on a template metric whose constant time slices are FLRW-like. The FLRW curvature parameter is assumed to be time-dependent in order to encode the effect of backreaction. Under these assumptions, in \cite{Larena:2008be} the authors perform a likelihood analysis of a scaling solution of the backreaction problem on their template metric, providing likelihood confidence contours of the present matter density parameter of the Universe ($\Omega_m^{D_0}$) and the scaling exponent $n$. The likelihood function is built from the SnIa of \cite{Astier:2005qq} and cosmic microwave background (CMB) data from \cite{Hinshaw:2006ia} and it is compared to a standard FLRW model with a DE with constant equation of state $w$.

In this paper we aim to put the most stringent constraints possible to the backreaction model proposed in \cite{Larena:2008be} with present background cosmological data. In particular, we consider a dataset which has never been used before in the context of backreaction: the set of angular diameter distances measured from Sunyaev Zeldovich (SZ) effect proposed by \cite{Bonamente:2005ct}. Unfortunately, it turns out that these data do not add relevant information due to the large errors associated with them.

Moreover, we update the SnIa and CMB data sets to the most recent catalogs: the \textit{Union2.1} SnIa catalog presented in \cite{Suzuki:2011hu}, the nine-year Wilkinson Microwave Anisotropy Probe  (\emph{WMAP}) data  \cite{Hinshaw:2012aka} and the \emph{Planck} \cite{Ade:2013zuv} data. The CMB data are used in two different ways: by exploiting the CMB shift parameters or by using the position of the peaks and dips of the temperature power spectrum.

The new SnIa dataset brings noticeable improvement to the constraints. Also the likelihood contours produced with recent CMB observations are quite tighter than those produced by older data.
Interestingly however, we find an inconsistency in the results produced by CMB data. We hence propose an explanation and possible ways out of this problem.

The paper is organized as follows. In Sec. \ref{sec:backreaction-theory} we summarize the features of the template metric proposed by \cite{Larena:2008be} in the context of backreaction, while in Sec. \ref{sec:data} we describe the data used to constrain it, i.e. clusters, SnIa and the CMB. Sec. \ref{sec:results} is devoted to the analysis of the data and to the presentation of the results obtained, with special emphasis on the CMB data in the context of backreaction. Conclusions are given in Sec. \ref{sec:conclusions}.

Natural units $c=1$ are assumed everywhere.

\section{Backreaction template metric}\label{sec:backreaction-theory}

Here we summarize the features and the main equations of the backreaction template metric introduced by \cite{Larena:2008be}, that we will use in our data analysis.

In \cite{Larena:2008be} the problem of testing averaged inhomogeneous cosmologies is handled by introducing a smoothed template metric. The template metric corresponds to a constant spatial curvature model at any time, but the curvature is enabled to evolve in time. It is introduced to describe the path of light in an averaged model and how it differs from the homogeneous counterpart. Since the true metric is unknown, the template introduces an approximated smoothed lightcone for the computation of observables.
The metric looks like \cite{Larena:2008be,Paranjape:2006ww}
\be
\label{template}
g_{\mu\nu}dx^{\mu}dx^{\nu}=-dt^2+\frac{1}{H_{D_0}^2}\frac{a_D^2}{a_{D_0}^2}\left(\frac{dr^2}{1-k_D(t)r^2}+d\Omega^2\right).
\ee 
Here the subscript $D$ indicates quantities averaged over a spatial domain $D$.
 The parameter $k_D(t)$, which describes the time variation of the 3-space curvature, cannot be arbitrary. It must be related to the full spacetime Ricci scalar, in analogy with the FLRW model. The factor $H_{D_0}^{-2}$, representing the Hubble function today, is inserted following \cite{Larena:2008be} in order to make the coordinate distance $r$ dimensionless.

Then, we need to assume that  the averaged Ricci scalar $\average{R}_D$ is related to $k_D$ as follows:
\be \label{eq:Ricciscalar}
\average{R}_D=\frac{k_D(t)|\average{R}_{D_0}|a^2_{D_0}}{a_D^2}.
\ee
Following \cite{Larena:2008be} let us compute the distance between two points on two slices at different cosmic time. First we remember that the distance between two points on the same slice is given by
\be \label{eq:distance}
l(t)=a_D(t)\int_0^r\frac{dx}{\sqrt{1-k_D(t)x^2}},
\ee
 and so we can compute the derivative of this distance with respect to cosmic time, in order to evaluate the infinitesimal distance between points belonging to different slices, separated by an infinitesimal time interval $dt$:
\be \label{eq:dist-sameslice}
\frac{dl}{dt}=H_D(t)l(t)+a_D(t)\frac{dk_D(t)}{dt}\int_0^r\frac{x^2dx}{1-k_D(t)x^2}.
\ee
We can identify the standard Hubble flow (first term on the right-hand side) and an extra term, which arises as a consequence of the time dependence of the curvature parameter. If we are interested on the path of a photon, the left-hand side of Eq.~(\ref{eq:dist-sameslice}) is simply the speed of light, so the equation reduces to
\be
\label{coordinate distance}
\frac{dr}{dt}=\frac{1}{a_D(t)}\sqrt{1-k_D(t)r^2},
\ee
 which is a differential equation for the dimensionless coordinate distance travelled by a photon.
In analogy with the standard model we introduce an effective redshift
\be \label{eq:effz}
1+z_D=\frac{(g_{ab}k^au^b)_S}{(g_{cd}k^cu^d)_O},
\ee
 where $k^a$ is the light wave-vector, $u^b$ is the comoving observer 4-velocity, S labels the source and O the observer. Since the wave-vector is normalized such that at the observer $k^au_a=-1$, we simply have
\be
1+z_D=(a_D k^0)_S.
\ee
The wave-vector satisfies the geodesic equation $k^{\nu}\nabla_{\nu} k^{\mu}=0$, which can be written in the context of the template metric as
\be
\label{geodesic equation}
\frac{1}{\hat{k}^0}\frac{d\hat{k}^0}{a_D}=-\frac{r^2(a_D)}{2(1-k_D(a_D)r^2(a_D))}\frac{dk_D(a_D)}{da_D}.
\ee
In this equation the function $r(a_D)$ is the dimensionless coordinate distance, which can be derived with respect to the effective scale factor by solving equation (\ref{coordinate distance}) written as
\be
\label{eq:coordinate distance second form}
\frac{dr(a_D)}{da_D}=-\frac{H_{D_0}}{a_D^2H_D(a_D)}\sqrt{1-k_D(a_D)r^2},
\ee
 with the boundary condition $r(z=0)=0$.
The previous model allows the computation of important observables in cosmology, such as the angular diameter distance $d_A$ and the luminosity  distance $d_L$:
\be \label{eq:lumdist}
d_A(z_D)=\frac{1}{H_{D_0}}a_D(z_D)r(z_D),  \quad
d_L(z_D)=(1+z_D)^2d_A(z_D).
\ee

As in \cite{Larena:2008be}, we consider solutions with exact scaling
\be \label{eq:scaling}
Q_D=Q_{D_i}a_D^p,  \qquad  \average{R}_D=\average{R}_{D_i}a_D^n,
\ee
 where the exponents are reals. 
 We are interested in the solution to the last constraint characterized by $n=p$,  that corresponds to a direct coupling between backreaction and the scalar curvature.
Inserting Eqs. (\ref{eq:scaling}) into Eq. (\ref{eq:Ricciscalar}) we find:
\be 
Q_D=-\frac{n+2}{n+6}\average{R}_{D_i}a_D^n,
\ee
and defining the $X$-component, accounting for the combined contributions of backreaction and curvature, as
\be
\Omega_X^D=-\frac{2\average{R}_{D_i}a_D^n}{3(n+6)H_D^2},
\ee
one can derive the equations that  will have to be solved in order to obtain $d_A$ and $d_L$:
\bea
&& k_D(a_D)=-\frac{(n+6)\Omega_X^{D_0}a_D^{n+2}}{|(n+6)\Omega_X^{D_0}|}, \\
&& H_D^2(a_D)=H_{D_0}^2(\Omega_m^{D_0}a_D^{-3}+\Omega_r^{D_0}a_D^{-4}+\Omega_X^{D_0}a_D^n),  \\
&& \frac{dr}{da_D}=\sqrt{\frac{1-k_D(a_D)r^2}{\Omega_m^{D_0}a_D^{-3}+\Omega_r^{D_0}a_D^{-4}+\Omega_X^{D_0}a_D^n}}.
\eea
The scaling of the $X$-component as $a_D^n$ shows that it is kinematically undistinguishable from DE in a FLRW cosmology.
 For completeness, we have generalized Larena's equations by adding  to the effective Hubble parameter the radiation component $\Omega_r^{D_0}$ at the given initial time, although the contribution of radiation is completely negligible.
We assumed $\Omega_r^{D_0}a_D^{-4}$ based on the analogy with the usual homogeneous case. In \cite{Buchert:2001sa} it is shown that the effective Friedmann equations are modified by the extra \textit{dynamical backreaction} term. This has the effect of smoothing out the inhomogeneities in the pressure field, but here we neglect the latter.
Larena et al. \cite{Larena:2006yk} showed that the effective inhomogeneous cosmology based on the averaging procedure could be written in a Friedmannian form if an effective energy-momentum tensor is defined. Its effective domain-dependent density and pressure are
\be
\rho^D_{eff}=\average{\rho}_D-\frac{1}{16\pi G}Q_D -\frac{1}{16 \pi G}\average{R}_D, \quad
p^D_{eff}=-\frac{1}{16\pi G}Q_D +\frac{1}{48 \pi G}\average{R}_D .
\ee
In terms of the effective sources,
\be
\label{conservation}
\dot{\rho}_{eff}^D+3H_D(\rho_{eff}^D+p_{eff}^D)=0.
\ee
In analogy with the standard homogeneous cosmology, we can write
\be
p_{eff}^D=w^D\rho_{eff}^D,
\ee
which implicitly defines a domain-dependent equation of state parameter $w^D$.
Comparing the scaling with $a$ of the $X$-component with the standard DE scaling, we immediately find a relation between $n$ and $w_D$:
\be
w^D=-\frac{n+3}{3}.
\ee

We stress that the backreacted model based on the template metric (\ref{template}) is different from a standard FLRW cosmology for two main reasons: first the coordinate distance satisfies a different differential equation, second the lightcone is slightly different. The two models are kinematically the same, since $H(a)$ and $H_D(a_D)$ have the same form, but the time evolution of the effective scale factor differs from the standard one due to the backreaction. In terms of redshifts, the two models are not equivalent: the non standard relation $a_D(z_D)(1+z_D)\neq 1$, which is a consequence of the non-standard lightcone associated to the template, clearly shows the difference.

We also stress that in order to build this backreaction model many assumptions have been made.  We summarize them here and refer to \cite{Larena:2008be}, where they were first introduced, for further detail.
\begin{itemize}

\item The template metric of Eq. (\ref{template}) has been used. This has been proposed by \cite{Larena:2008be} justified by the fact that Ricci flow renormalization of the average characteristics on an inhomogeneous space-time decreases intrinsic curvature inhomogeneities and produces at any given time a constant curvature slice, but at different times these constant curvatures are not equal. Based on different considerations, also \cite{Paranjape:2006ww} propose the same metric.

\item The relation between the curvature $k_D(t)$ and the Ricci scalar of Eq. (\ref{eq:Ricciscalar}) is also an assumption, again proposed by \cite{Larena:2008be} in analogy with the FLRW metric, to which it reduces when posing $k_D (t_0) |\average{R}_{D_0}| = k_{D_0}/6$, $k_D(t_0) = 1$ on a large domain $D_0$. Eqs. (\ref{eq:distance}-\ref{coordinate distance}) are consequences of these two assumptions.

\item The effective redshift of Eq. (\ref{eq:effz}) is defined in analogy to FLRW, to which it reduces when the effective metric becomes FLRW.

\item The reciprocity relation Eq. (\ref{eq:lumdist}) is implicitly assumed to work in the averaged spacetime. This assumption, although reasonable, is somewhat critical: on one side it follows from the metricity of spacetime only, which leads to believe that it should correctly hold for the template; on the other side however, the coarse-graining procedure of spacetime changes the lightcone and the cross-sectional areas subtended to ray bundles should change. In summary, whether the Etherington relation is safe to use in the averaged framework, is not certain.

\item As \cite{Larena:2008be}, we restrict to the scaling solutions shown in Eq. (\ref{eq:scaling}), and only to the case $n=p$, while other solutions may exist.

\item Finally, as again stated in \cite{Larena:2008be}, a strong albeit reasonable assumption is that at early times the spacetime is described by a weakly perturbed FLRW. This leads to the definition of the sound horizon and of the fundamental multipole that can be found in Sec. \ref{sec:cmb}. This assumption is reasonable because the backreaction effect should be very small as inhomogeneities get smaller. On the other hand it is easy to see that the template metric, which is not perturbative, would not lead to the same definition of the sound horizon if it is assumed as the underlying spacetime metric of the primordial Universe as soon as $k_D(t)$ is not constant. Moreover, to be precise, by assuming that the fundamental multipole is as in the standard CMB theory we state that the effect of backreaction upon the observed spectrum is completely accounted in a different angular size subtended at the decoupling epoch.

\end{itemize} 

\section{Data sets} \label{sec:data}
In this section, we describe the data sets used to constrain the template backreaction metric illustrated in the previous section. These are SZ clusters, SnIa and the CMB.

\subsection{SZ clusters}\label{sec:szclusters}

It is well known that the SZ effect allows the detection of galaxy clusters \cite{Zeldovich:1969en,Sunyaev:1970eu,Sunyaev:1970er} through the following mechanism. The gas of the intra-cluster medium (ICM) shifts the frequency of photons coming from the last scattering surface through inverse Compton scattering. This frequency shift alters the CMB temperature fluctuation and can be detected.
The SZ effect allows a computation of $d_A$ that intrinsically depends on the physics of the ICM and the cluster geometry, whereas it is independent of the background cosmology.
This is done through the combined measurement of the bolometric surface brightness  
of the cluster in the X-ray band and of the SZ intensity fluctuations.
The angular diameter distance of a cluster is given in terms of these:
\be \label{eq:dA-sz}
d_A=\frac{N^2_{SZ}}{N_X} \left( \frac{\Lambda_e^{(0)}}{4\pi(1+z)^4 I_0^2\psi_0^2\sigma_T^2} \right).
\ee
Here $N_{SZ}$ and $N_X$ are quantities proportional to the bolometric surface brightness and to the SZ intensity fluctuations, respectively, $I_0$ is the reference intensity $I_0= (2h/c^2)(k_B T/h)^3$,
and $\sigma_T$ is the Thompson cross-section.  
$\Lambda_e^{(0)}$ and $\psi^{(0)}$ are parameters related to the electron cooling function and to the relativistic corrections to the inverse Compton scattering, respectively.
Note that in Eq. (\ref{eq:dA-sz}) the factor $(1+z)^{-4}$ follows directly from the ray-tracing equation in the usual FLRW, which provides the well-known relation between the redshift and the scale factor $1+z= a(t)/a(t_0)$.
Since in the template metric the curvature $k_D$ is time-dependent, the standard ray-tracing equation is substituted by Eq. (\ref{geodesic equation}) and the previous relation does not hold anymore. For simplicity we assume that Eq. (\ref{eq:dA-sz}) can be safely used in the backreacted model, provided that the redshift is substituted by the effective redshift.

The SZ clusters catalog we use has been compiled by Bonamente et al. \cite{Bonamente:2005ct} and it is composed of 38 clusters in the redshift range $0.14 \ < \ z \ < \ 0.89$. The distances are computed combining \textit{Chandra} X-ray data and SZ measures from the \textit{Owens Valley Radio Observatory} (OVRO) and the \textit{Berkeley-Illinois-Maryland Association} (BIMA) interferometric arrays. 

As we see from Eq. (\ref{eq:dA-sz}), $d_A$ depends also on the geometry of the cluster through the ratio $N_{SZ}^2/N_X$. This depends on the physical properties of the ICM. Many models for the ICM were proposed in the past years. 
In the Bonamente et al.  catalog, the ICM is modeled using both the \textit{double} $\beta$-model and the spherical $\beta$-model.
The spherical $\beta$-model \cite{Cavaliere:1976tx} is an isothermal 
 model based on the spherical simmetry of the electron density.
Instead, the \textit{double} $\beta$-model of Mohr \cite{Mohr:1999ya} and Laroque \cite{Laroque:2005ws} is based on a weightened superposition of two copies of the isothermal $\beta$-model  (see \cite{LaRoque:2006te,Bonamente:2005ct} for a comparison between the models).

As regards the errors, we follow \cite{Chen:2011na} who use the same cluster catalog to constrain DE. There are three different contributions for the error $\sigma_{data}$: $\sigma^2_{data}=\sigma^2_{stat}+\sigma^2_{sys}+\sigma^2_{mod}$,
 where $\sigma^2_{stat}$ and $\sigma^2_{sys}$ are the statistical and systematic uncertainties respectively, and $\sigma_{mod}$ accounts for the error in modelling the cluster. The modelling error are given in percentage in Table I of \cite{Bonamente:2005ct}, while $\sigma^2_{stat}$ and $\sigma^2_{sys}$ are listed in Table III. 

Since the systematic errors are asymmetric, also the total error is such, but in our analysis we take a symmetric error bar, using the larger error among the two associated to each measurement, in order to be conservative. 
$d_A$ depends parametrically on $H_0$, which we treat as a nuisance parameter and marginalize over, assuming a Gaussian prior centered in $H_0=72 $ Km/s/Mpc with $\sigma_{H_0}=8$ Km/s/Mpc \cite{Riess:2011yx}. 
We test the dependence of our results on the prior by trying a flat prior on a wide interval: $H_0\in (0,200)$. It turns out that this does not affect noticeably the probability contours, showing that contributions to the marginalized likelihood coming from higher or lower values of $H_0$ are intrinsically small, regardless of how they are weighted by the prior.

\subsection{Supernovae Ia}

In order to compare our constraints with those obtained in \cite{Larena:2008be}, we first perform our analysis on the \textit{Supernova Legacy Survey} (\emph{SNLS}) catalog  \cite{Astier:2005qq}. This sample is composed by $115$ SnIa in the redshift range $0.249 <z <1.01$.
We then update the analysis by using a second set of SnIa: the \textit{Union2.1} sample (see \cite{Suzuki:2011hu}), which is an update of the \textit{Union2} sample \cite{Amanullah:2010vv}. The \textit{Union2.1}  
sample is presently one of the largest SnIa set reaching high redshift, and it is composed of $580$ SnIa with redshifts ranging from $z=0.015$ to $z=1.4$.
The SnIa observable is the distance modulus, related to $d_L(z,\theta)$ through 
\be
\mu(z)=5\log_{10}(H_0 d_L)+\mu_0,
\ee
where $\mu_0$ is a nuisance parameter which encodes the dependence on the SnIa absolute magnitude $M$ and of $H_0$, and over which we marginalize analytically, and  the luminosity distance $d_L$ can be written in the context of a flat FLRW Universe as 
\be \label{eq:lum-dist}
d_L=(1+z)\int_0^z\frac{1}{H(z')}dz'.
\ee
As regards the errors of the \emph{Union2.1} catalog,
in our work we neglect the covariance contribution (see \cite{Kowalski:2008ez,Amanullah:2010vv,Suzuki:2011hu}), as done e.g. by \cite{Lazkoz:2007zk,Sendra:2011pt}.
Also, instead of using Eq. (\ref{eq:lum-dist}), we find it computationally less demanding to compute the comoving distance $r_{cm}$ by directly solving the differential equation
\be \label{eq:com-dist_flrw}
\frac{dr_{cm}(a)}{da}=\frac{\sqrt{1-kr_{cm}^2}}{H(a)a^2}.
\ee
in the case of FLRW, while in the case of backreaction $r_{cm} = r/H_{D_0}$, where $r$  is the solution of Eq. (\ref{eq:coordinate distance second form}).

\subsection{CMB} \label{sec:cmb}

We make use of two different sets of CMB parameters. These are the CMB shift parameters and the location of the peaks and dips in the temperature power spectrum.

\subsubsection{Acoustic peaks}\label{sec:acoustic-peaks}

The computation of the power spectrum of the CMB requires to solve the Boltzmann equation for the photon-baryon fluid and the Einstein field equations. A full solution has been derived in the context of a linearly perturbed FLRW spacetime. 
Since the inhomogeneities are negligible at the recombination epoch, it is assumed that the early Universe can be considered as a perturbed FLRW up to recombination, and that the standard solution is still valid in a backreacted scenario. Hence, the location of the CMB peaks can be computed without any assumption on the late-time cosmology.

The position of the $m-th$ multipole is given by
\be 
l_m=l_a(m-\phi_m),
\ee
 where $m$ takes positive integer values for the peaks, and half-integer positive values for the dips. The correction term $\phi_m$ depends on the matter content of the early Universe $\Omega_m h^2$, on the baryon density $\Omega_b h^2$, on the redshift of last scattering $z^*$, and on the spectral index $n_s$.  These parameterize the effect of gravitational dragging occurring up to recombination, which shifts the position of the peaks with respect to the characteristic scale given by $r_s$. To compute the position of each peak we use the fitting formulae given by Doran and Lilley in \cite{Doran:2001yw}. We neglect DE, because its amount is negligible at the redshift of recombination in standard DE models. We set the spectral index to the best fit value measured by \emph{Planck} $n_s=0.96$ \cite{Ade:2013uln}, instead of setting it to $n_s=1$ as in \cite{Larena:2008be}. We let $\Omega_b h^2$ vary, together with the other parameters $\Omega_m$, $w$ and $H_0$. 
The fundamental multipole $l_a$ is defined as
\be
\label{CMB multipole}
l_a=\pi \frac{r_{cm}(a^*)}{r_s(a^*)},
\ee
 where $a^*$ is the scale factor corresponding to the epoch of recombination, $r_{cm}$ is the comoving distance and $r_s$ is the comoving sound horizon, which is defined as 
\be
\label{sound horizon}
r_s(a^*)=\int_0^{a^*} c_s d\eta,
\ee
 where $c_s$ is the speed of the acoustic waves propagating through the primordial photon-baryon plasma and $\eta$ is the conformal time defined through $d\eta=dt/a(t)$.
The sound speed encodes the microphysics of the photon-baryon fluid and it depends on the ratio between the baryon and the photon content of the plasma:
\be 
\label{speed of sound}
c_s=\frac{1}{\sqrt{3(1+R(a))}},
\ee
 where
\be
\label{ratio}
R(a)=\frac{3\rho_b(a)}{4\rho_{\gamma}(a)}=\frac{3\Omega_b}{4\Omega_{\gamma}}a
\ee
Here $\Omega_b$ and $\Omega_{\gamma}$ are the baryon and the photon density parameter today, respectively, while $\rho_b$ and $\rho_{\gamma}$ are the corresponding densities. Other relativistic species (like neutrinos) do not enter this ratio, but they are involved in the computation of the comoving distance. 

Combining Eqs. (\ref{sound horizon}-\ref{ratio}) and recalling that the
 Hubble function for a flat FLRW Universe is given by
\be
H(a)=H_0\sqrt{\Omega_m^0a^{-3}+\Omega_r^0a^{-4}+(1-\Omega_m^0-\Omega_r^0)a^{-3(1+w)}},
\ee
 we can compute the sound horizon from the integral
\be
r_s(a^*)=\frac{1}{H_0}\int_0^{a^*}\frac{da}{a^2\sqrt{\Omega_m^0 a^{-3}+\Omega_r^0 a^{-4}+(1-\Omega_m^0-\Omega_r^0)a^{-3(1+w)}}\sqrt{3(1+R(a))}}.
\ee 
The radiation density $\rho_r$ is given by
\cite{Hinshaw:2012aka}
\be
\rho_r=\rho_{\gamma}\left[1+\frac{7}{8}\left(\frac{4}{11}\right)^{4/3}N_{eff}\right]\approx \rho_{\gamma}(1+0.2271N_{eff}),
\ee
where $N_{eff}$ is the effective number of neutrino species, which takes into account the possible existence of extra relativistic species. In this work we  assume the standard model value $N_{eff}=3.04$, for simplicity, and because observations are consistent with this value \cite{Ade:2013zuv}.
The scale factor at recombination $a^*$  is computed in terms of the corresponding redshift $z^*$  through the standard relation $a^*=1/(1+z^*)$, where  $z^*$ is determined through the fitting formulae given in \cite{Durrer:2001jz}, \cite{Hu:1995en} and references therein.

Finally, the correction terms $\phi_m$ to the position of the peaks are given by the fitting formulae of Doran and Lilley in \cite{Doran:2001yw} and depend on the ratio
\be
r^*=\frac{\Omega_r(z^*)}{\Omega_m(z^*)}.
\ee  
In \cite{Larena:2008be}  the positions of the CMB peaks from \emph{WMAP-3} data \cite{Hinshaw:2006ia} were used. To update their analysis, we use the more recent \emph{WMAP-9} and \emph{Planck} data, but since
no measurement of the position of the peaks of the power spectrum of the CMB were available for these datasets at the time of preparing this work, we had to compute them.
Our method is similar to that of \cite{Page:2003fa}, used also by \cite{Hinshaw:2006ia} for \emph{WMAP-3}: we fit the peaks and dips with exponential functions and parabolas, but differently from it, we fix the boundaries delimiting each peak instead of treating them as free parameters. We used this approach for simplicity and time economy, because a full fitting procedure  would require at least 14 parameters in order to reconstruct the shape of the spectrum up to the third peak.  For a detailed explanation of the  fitting procedure see Appendix \ref{app:peaks}. In Table \ref{Tabella picchi} we  list the position of the first three peaks and of the first dip of the CMB spectrum, together with the corresponding error. 
\begin{table}[ht]
\caption{Positions of the CMB peaks and related errors}
\centering
\begin{tabular}{cccccc}
\hline\hline
     & \vline   &  WMAP-9  & \vline &   Planck  \\
\hline
$l_1$   & \vline &   $220.9\pm 0.9$  & \vline & $219.9 \pm 0.7$  \\
$l_{dip}$ & \vline  & $415.4 \pm 1.7$  &  \vline & $419.2 \pm 1.1$  \\
$l_2$ & \vline  &   $537.8 \pm 2.9$  &  \vline & $537.0 \pm 1.7$  \\
$l_3$  & \vline  &  $813.5 \pm 9.8$  &  \vline & $813.6 \pm 1.6$ \\
\hline
\end{tabular} 
\label{Tabella picchi}
\end{table}
We see that the positions of the CMB peaks obtained by fitting \emph{WMAP-9} data are in excellent agreement with those provided by \emph{Planck}. The position of the first dip, on the other hand, shows only marginal consistency among the two data sets. A rough estimate of the deviation gives
\be
|l_{dip}^{Planck}-l_{dip}^{WMAP}|\sim 2.8\bar{\sigma}
\ee
 where $\bar{\sigma}=(\sigma_{WMAP}+\sigma_{Planck})/2$. 
The reason for this can be found in our fitting procedure. As said previously, we arbitrarily fix the boundaries delimiting each peak. In the case of the dip, this is problematic because it is a narrow feature with respect to the peaks and  few data points, in the case of \emph{Planck} data, are used to determine its position (see bottom panel of Fig. \ref{fig:spectra-fits}). If we had treated the boundaries of the dip as free parameters, our error on its position would likely increase enough to make the \emph{Planck} measurement compatible with the \emph{WMAP-9} one.
For more details see the Appendix \ref{app:peaks}.

When analyzing the data, we marginalize numerically over $\Omega_bh^2$ assuming a Gaussian prior $\Omega_bh^2=0.0214 \pm 0.0020$ coming from Big Bang Nucleosynthesis data \cite{Kirkman:2003uv}, and  over $H_0$ with the same prior used for clusters data (see Sec. \ref{sec:szclusters}).

\subsubsection{CMB shift parameters}

The shift of the acoustic peaks of the CMB can be quantified by using a different set of observables, called CMB shift parameters: $(l_a,R,z^*)$ \cite{Wang:2007mza,Nesseris:2010pc}. Here $l_a$ is the fundamental multipole of the CMB defined in (\ref{CMB multipole}), $z^*$ is the redshift of recombination and $R$ is defined by~\cite{Efstathiou:1998xx}
\be \label{eq:shift-1}
R=\sqrt{\Omega_m^0}H_0\int_0^{z^*}\frac{dz}{H(z)}
\ee
 for a flat FLRW Universe. We use the CMB shift parameters computed from \emph{WMAP-9} data in \cite{Hinshaw:2012aka}. 
Again, as done for the luminosity distance, instead of computing the integral in Eq. (\ref{eq:shift-1}), that corresponds to the comoving distance, we compute it by directly solving  Eq. (\ref{eq:com-dist_flrw}) for the FLRW case, and, for backreaction, by solving Eq. (\ref{eq:coordinate distance second form}), thus obtaining the dimensionless $r$, related to the comoving distance by $r_{cm} = r/H_{D_0}$. 
The parameter $R$ has been used to analyze CMB data many times in the context of FLRW models (see for instance \cite{Komatsu:2008hk,Elgaroy:2007bv,Wang:2007mza,Corasaniti:2007rf}) and in local voids ones \cite{Clifton:2009kx}. The model dependency of $R$ has been discussed by several authors. For example, in \cite{Durrer:2001jz} the authors suggest a way for extracting model-independent constraints from it. In \cite{Corasaniti:2007rf} instead,  a likelihood analysis with \emph{WMAP-3} data fot $R$ and $l_a$ is carried out involving extra parameters, like positive neutrino masses and tensor modes. They summarize their results in their Table 1, which shows that  changing cosmic curvature or slightly modifying the DE parameters does not significantly change the value of $R$. On the other hand, the dependency on more exotic parameters, like non-zero neutrino masses, tensorial modes or a running spectral index, is much stronger. Under our assumptions, the $R$ parameter should be stable, since no tensorial modes nor massive neutrinos are considered. On the other hand, although $R$ is almost the same for many DE models, this parameter must be used with caution. The method based on the position of the peaks described in Sec. \ref{sec:acoustic-peaks} has a wider field of application, as both \cite{Corasaniti:2007rf} and \cite{Doran:2001yw} state, showing that it still works well both for models with early DE and models with late-time geometry or photon dynamics departing from a standard FLRW. The latter case is indeed relevant for our work.

\section{Data analysis  and results} \label{sec:results}

We analyze the different data sets separately using the backeaction template model and a FLRW Universe with DE with constant equation of state, and then we compare the results obtained for the two different cosmologies.

In the following, we used the notation $\Omega_m^{D_0}$ everywhere (instead of using the more standard $\Omega_m$ for the FLRW case).

\subsection{FLRW cosmology}

\begin{figure}[!th]
\begin{center}
\includegraphics[width=0.4\textwidth]{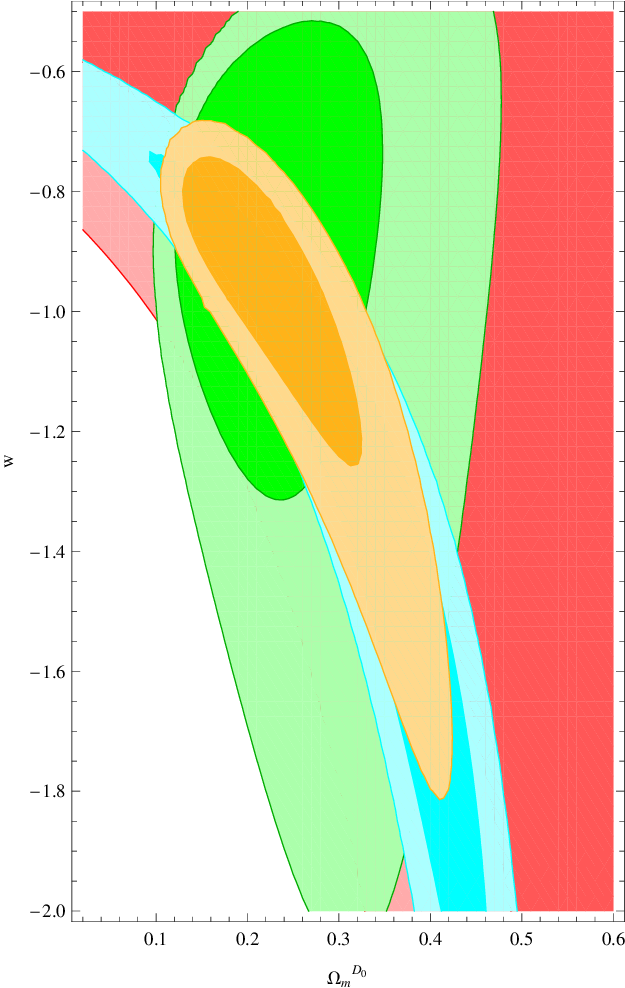}\quad
\includegraphics[width=0.4\textwidth]{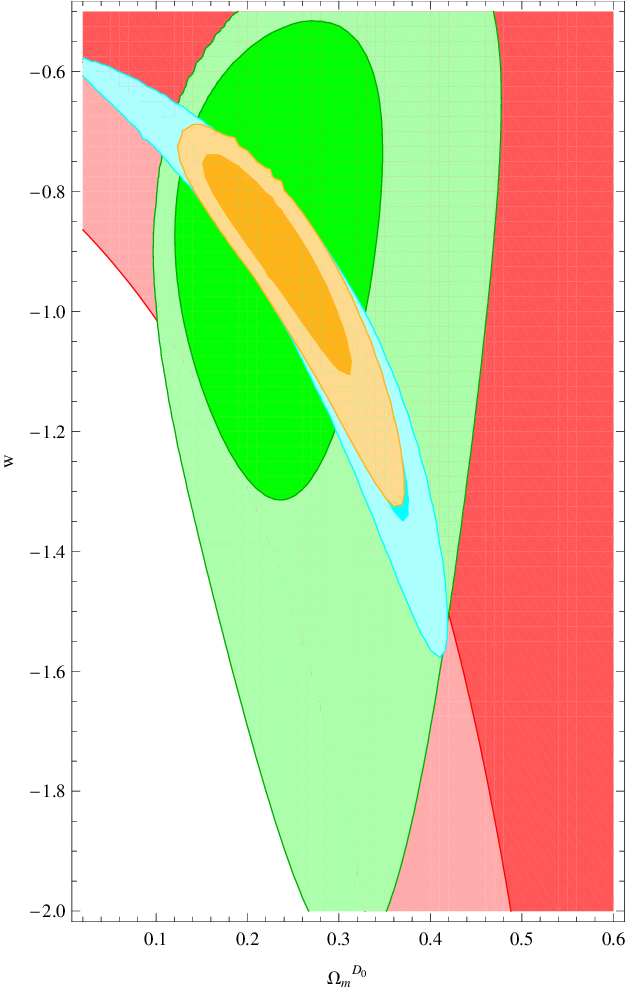}
\includegraphics[width=0.4\textwidth]{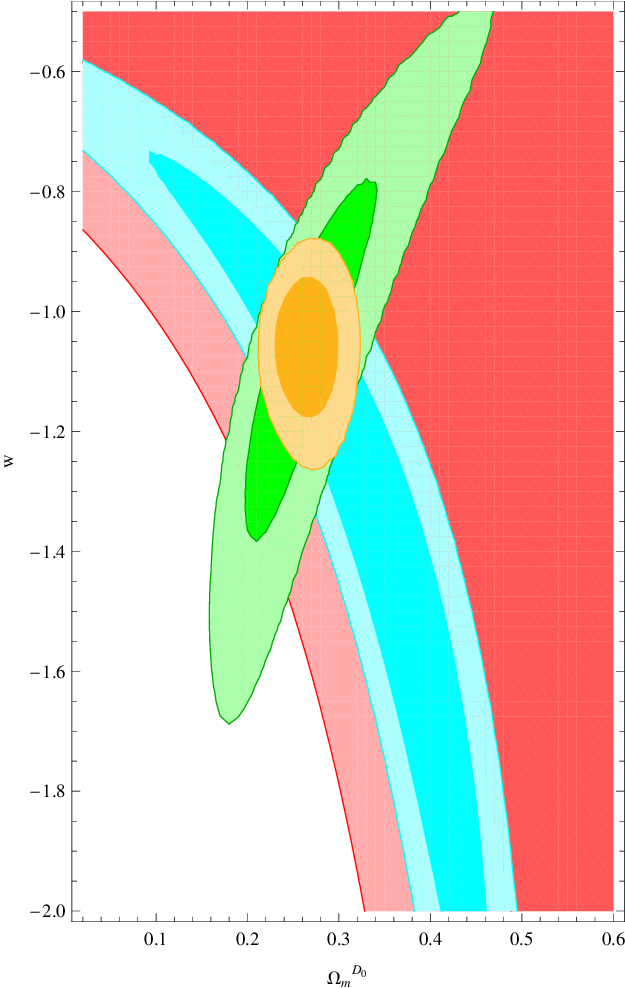}\quad
\includegraphics[width=0.4\textwidth]{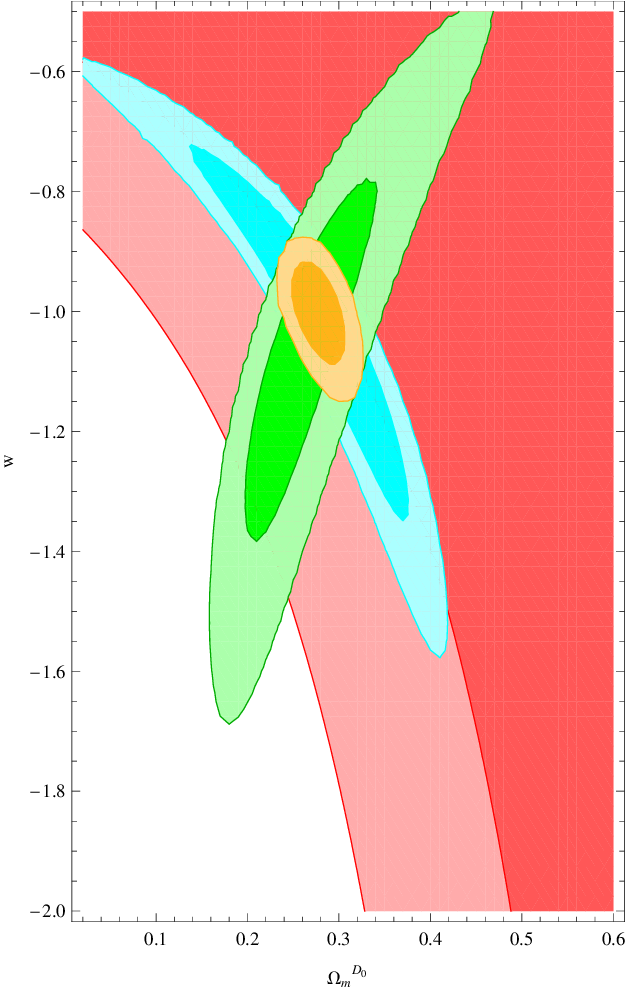}
\caption{$1\sigma$ and $2\sigma$ likelihood contours for the effective parameters $\Omega_m^{D_0}$ and $w$ for the FLRW model. 
Red fields are given by clusters. Green fields comes from CMB data: \emph{Planck} constraints from CMB peaks are given in the upper panels, \emph{WMAP-9} constraints from the CMB shift on the lower ones. The blue fields correspond to SnIa: \emph{SNLS} constraints are shown in the left panels, \textit{Union2.1} constraints are in the right panels. Joint constraints are shown in yellow.}
\label{fig:FLRW-summary}
\end{center}
\end{figure}

Here we present the analysis of the constraints on the flat FLRW, 
whose most important aspects are summarized in Fig. \ref{fig:FLRW-summary}.

The red regions correspond to constraints coming from clusters data. These are unfortunately very wide due to the huge uncertainties in modelling the cluster geometry. By comparing the red fields with the blue ones, corresponding to constraints from SnIa, it is clear that the latter are fully contained in the former. This means that present cluster data cannot significantly improve the constraints we can infer from SnIa. It is nevertheless interesting to notice that these two completely independent datasets are fully consistent with each other in the context of our FLRW cosmology.

If we compare the blue fields of the left panels, corresponding to the \emph{SNLS} SnIa data, with those on the right panel, corresponding to the \textit{Union2.1} data, we notice that the improvement in the SnIa constraint is significant, due to the much larger number of data of the latter sample.

The green fields indicate constraints from CMB data. The upper panels  show constraints given by the position of the first three peaks and the first dip of the CMB power spectrum from the \emph{Planck} data. The bottom ones show constraints derived from the method based on the CMB shift parameters computed from the \emph{WMAP-9} data. Since the procedure to fit the position of the CMB peaks is not very refined, the first dataset produces worse constraints with respect to the second one (see Appendix \ref{app:peaks} for more details).
We do not show constraints derived from the position of the peaks from  \emph{WMAP-9} data, extracted with the same method used for the \emph{Planck} data, because the errors we obtained were larger than those from \emph{Planck}.

Overall, the most important result shown in Fig. \ref{fig:FLRW-summary} is that all combinations of data are fully consistent with the $\Lambda$CDM model, which is represented by a constant line $w=-1$, as found in most recent work (see e.g. \cite{Samushia:2013yga} and references therein).

\begin{figure}[!ht]
\begin{center}
\includegraphics[width=0.4\textwidth]{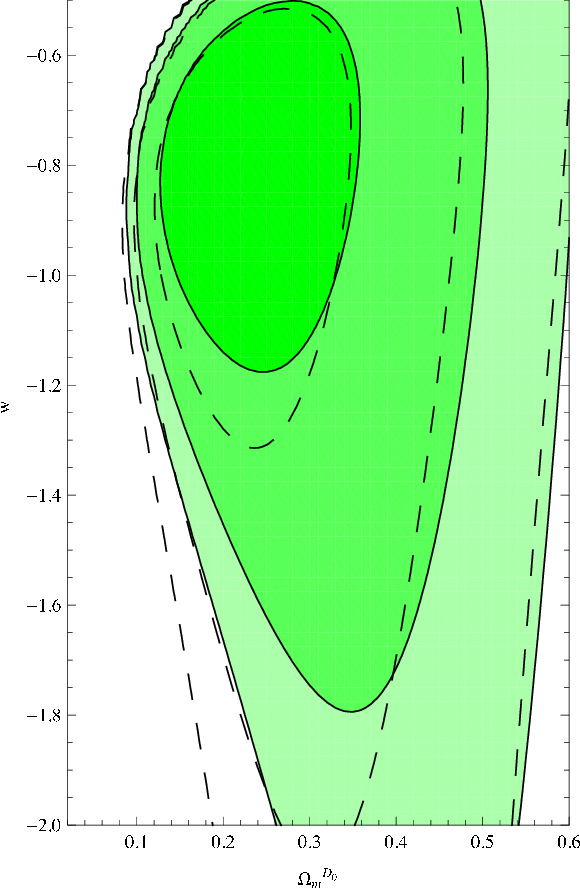}
\includegraphics[width=0.4\textwidth]{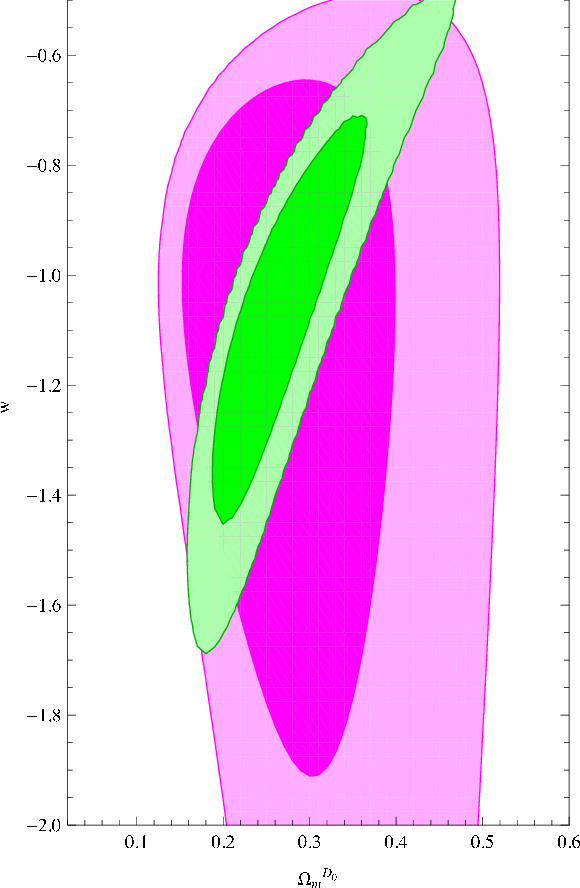}
\caption{$1\sigma$, $2\sigma$ and $3\sigma$ likelihood contours for the effective parameters $\Omega_m^{D_0}$ and $w$ for the FLRW model. Left panel: improvement of \emph{Planck} constraints (empty contours) combining clusters data (filled contours). Right panel: comparison between constraints provided by the position of the CMB peaks and dips (purple fields) and those given by the CMB shift parameters (green fields); only \emph{WMAP-9} data were used.}
\label{fig:clusters-improvement}
\end{center}
\end{figure}

For completeness, in Fig.  \ref{fig:clusters-improvement} we treat two interesting aspects. The left panel shows the improvement induced on the constraints given by \emph{Planck}  when combining them with constraints from clusters data. Even if combined constraints do not present any significant improvement, clusters data slightly move \emph{Planck} constraints  towards a larger value of $\Omega_m^{D_0}$, suggesting that more precise measurements of cluster distances based on the SZ-effect may be used as an independent probe of cosmology which can be fruitfully combined with CMB data. 
The right panel of Fig.  \ref{fig:clusters-improvement} shows a comparison between constraints coming from two different sets of \emph{WMAP-9} data. The wide purple fields correspond to constraints from the position of the first three peaks and the first dip of the CMB power spectrum, the green contours are constraints derived from the CMB shift parameters. The purple fields are very wide due to larger errors of \emph{WMAP-9} and to our fitting procedure. The method based on the shift parameters gives tighter constraints and predictions from \emph{Planck} data are more precise. 
We also note that the constraints derived from the shift parameters in the right panel of Fig.  \ref{fig:clusters-improvement} are completely contained into the constraints coming from the position of the peaks.
This is because both sets of parameters depend on $l_a$, but it seems that the shift parameters encode more information included in the CMB spectrum.

\subsection{Backreaction cosmology}

Here we discuss our results concerning the backreaction model based on the template metric of Eq. (\ref{template}).
Our results are compared with those summarized by Fig. 2 of \cite{Larena:2008be}. Here the authors show combined constraints from the SnIa of \cite{Astier:2005qq} and the positions of the CMB peaks from $WMAP$-3. The main feature is clear: likelihood contours move slightly towards a  Universe with larger $\Omega_m^{D_0}$ if backreaction is assumed. This effect can be considered the signature of curvature: the bare metric is curved and any \textit{on average} description of the Universe can be separated from the corresponding flat FLRW cosmology since the coarse-graining procedure does not completely destroy the information about the curvature of the bare metric. Results are shown in the plane $\Omega_m^{D_0}$-$n$, while we show constraints on $\Omega_m^{D_0}$-$w$, but it is easy to relate the two planes, knowing that $w = -(n+3)/3$.

\begin{figure}[!ht]
\begin{center}
\includegraphics[width=0.3\textwidth]{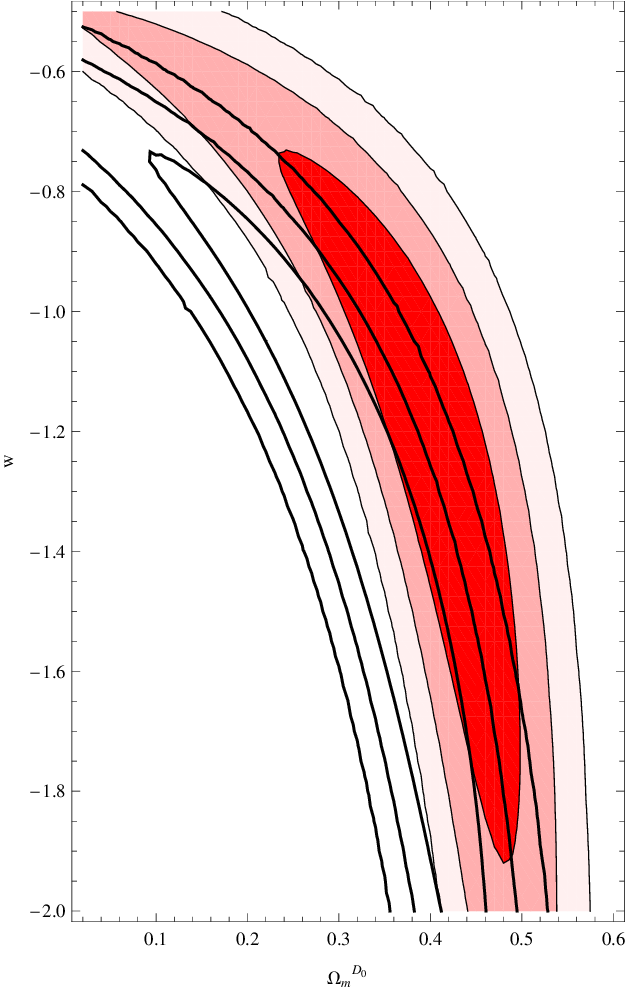}
\includegraphics[width=0.3\textwidth]{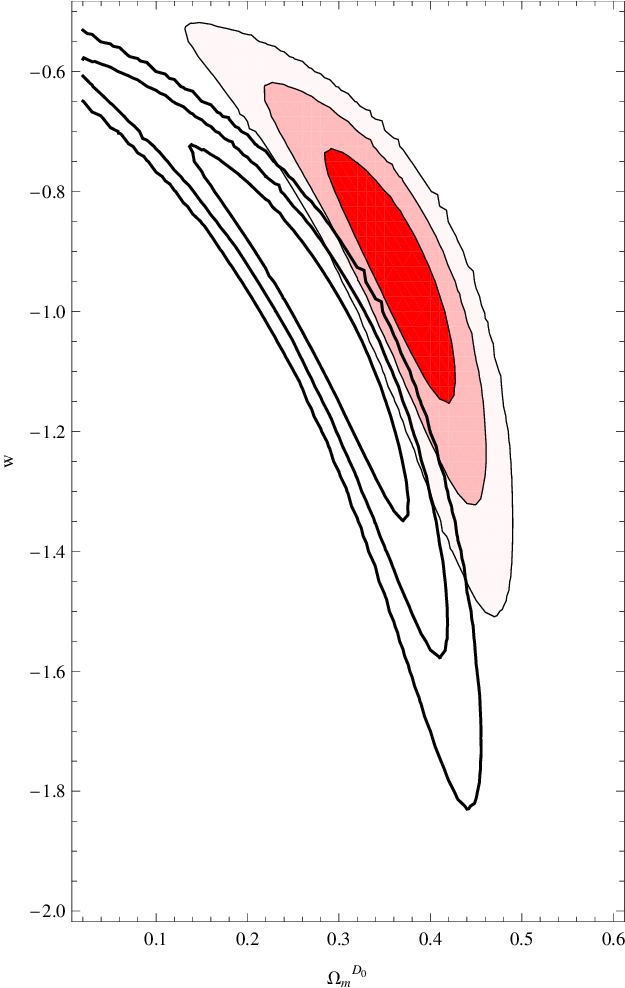}
\includegraphics[width=0.3\textwidth]{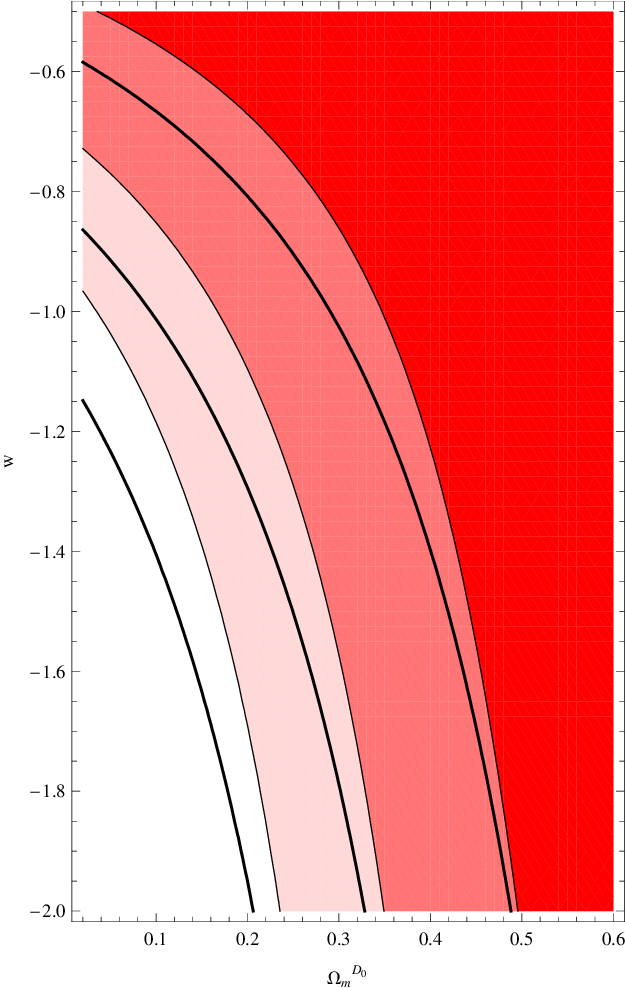}
\includegraphics[width=0.3\textwidth]{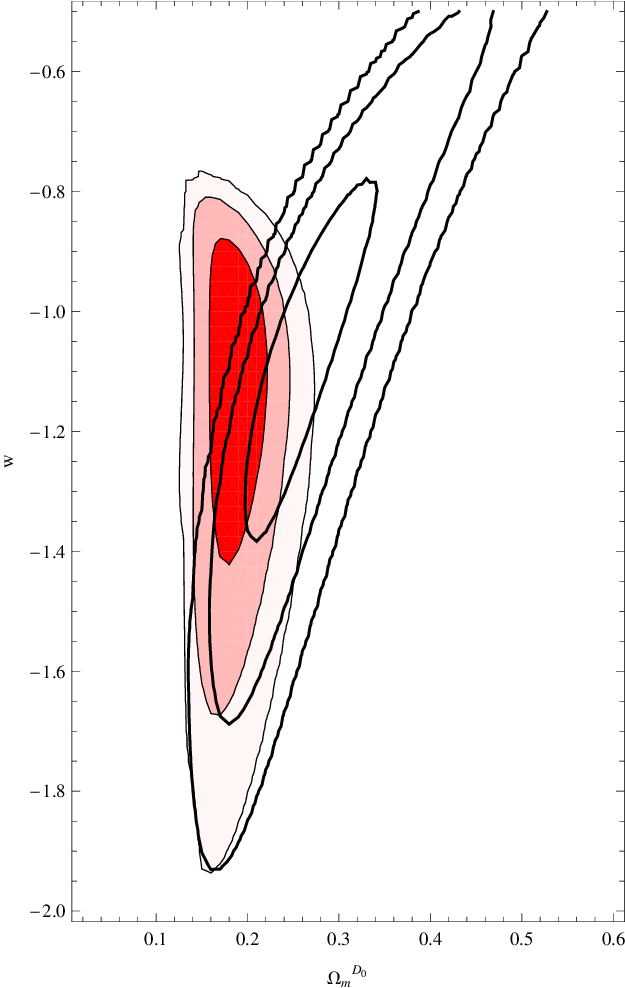}
\includegraphics[width=0.3\textwidth]{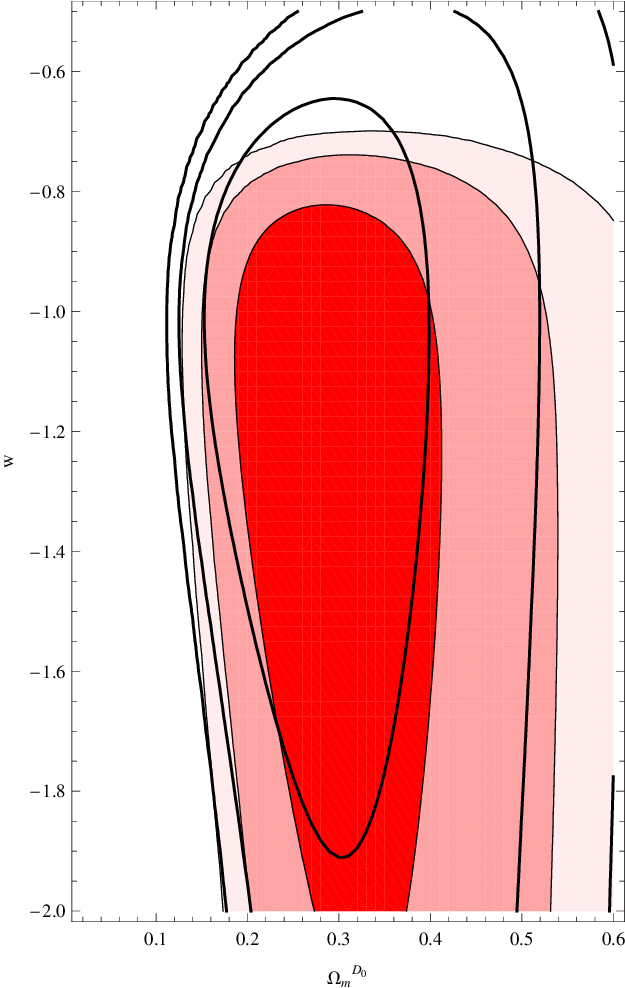}
\includegraphics[width=0.3\textwidth]{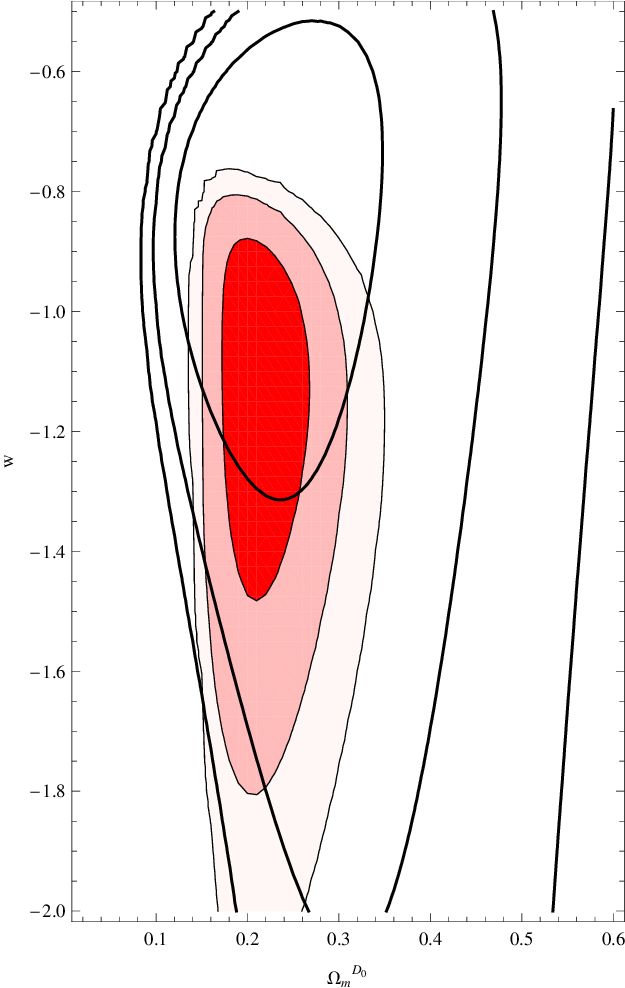}
\caption{Comparison between the backreaction model (filled contours) and a flat FLRW with DE with constant equation of state (blank contours). $1\sigma$, $2\sigma$ and $3\sigma$ likelihood contours are shown for different data sets. Upper left: \emph{SNLS} data. Upper central: \textit{Union2.1} data. Upper right: Clusters. Lower left: \emph{WMAP-9}  CMB shift parameters. Lower central: \emph{WMAP-9}  CMB peaks position. Lower right: \emph{Planck} CMB peaks position.}
\label{fig:backreaction-summary}
\end{center}
\end{figure}

In Fig.  \ref{fig:backreaction-summary} we show results coming from each dataset separately. Contours in the upper left panel comes from the \emph{SNLS} data and show the behavior we expected: we recover the results shown in Fig. 2 of \cite{Larena:2008be}. Indeed, filled regions, corresponding to the backreaction model, are slightly pushed towards higher values of $\Omega_m^{D_0}$. The \textit{Union2.1} catalog (see the upper central panel) behaves likewise. Constraints from this data set are tighter than those given by the \emph{SNLS}. This is mainly due to the larger number of data points of the \emph{Union2.1} ($580$ vs $117$). We note for this dataset that contours corresponding to the backreacted scenario overlap those for the flat FLRW model only at $3\sigma$, which suggests that SnIa observations may provide in the future a way for distinguishing between FLRW models and backreacted ones, if combined with other observations.

The upper right panel of Fig.~\ref{fig:backreaction-summary} shows constraints from clusters data. The big errors on the cluster geometry project onto very large errors on  $\Omega_m^{D_0}$ and $w$, so that constraints are very loose, but we observe that contours relative to backreaction have the expected behavior of preferring a higher value of $\Omega_m^{D_0}$, as happens for SnIa.

In the lower panels of Fig.~\ref{fig:backreaction-summary} we show constraints derived from our sets of CMB data.  We note that the positions of the CMB peaks extracted with our fitting procedure give worse constraints with respect to CMB shift parameters.
We also notice that the left edges of the probability contours are sharply cut in the direction of constant $\Omega_m^{D_0}$. We verified that this is due to the choice of the flat prior on $H_0$, and it happens both for FLRW and for backreaction. 

More importantly, we see that \emph{Planck}'s peak positions and \emph{WMAP-9} CMB shift parameters, that give relatively small errors, give contours that move towards lower values of $\Omega_m^{D_0}$, contrarily to what happens for SnIa and  clusters, and to what expected in general for backreaction models. (In the case of \emph{WMAP-9}, constraints derived from the positions of the peaks on the CMB power spectrum are very loose and the $1\sigma$ regions corresponding to the backreacted model and to the FLRW overlap.)
We note that this effect regards both the CMB shift constraints and the peaks positions ones, when considering backreaction. Since the CMB shift parameters are slightly model-dependent, we could have ascribed to this problem the strange behavior of such constraints, moving towards smaller values of $\Omega_m^{D_0}$, but the fact that this also happens for the peaks positions, that are less model dependent, ensures that such hypothesis is wrong.
We propose instead the following explanation.
Let us suppose that the estimate of the recombination redshift $z^*$ from Doran and Lilley's fitting formulae \cite{Doran:2001yw} is valid both in the case of a FLRW  and of a backreaction cosmology, since it depends only on the early time cosmology, which to a good approximation is the same in the two different models. This seems to be a reasonable assumption, but it can cause a theoretical problem, as we will explain in the following.
The problem arises when computing the fundamental multipole $l_a = l_a(a_D)$. As $l_a$ depends on $a_D$, one has to compute $a_D$ from $z^*$. To do so in our backreaction model, in principle  we have to use Eq. (\ref{coordinate distance}), as the relation $a=1/(1+z)$ does not hold in the backreaction model.
Using such correspondence, the authors of \cite{Rosenthal:2008ic} found that their model of backreaction was inconsistent with CMB data due to a wrong prediction of the size of the sound horizon. We found the same inconsistency between our template model and the CMB data of \emph{WMAP-9} and \emph{Planck}.
This is in contrast to what happens in  \cite{Larena:2008be}, where the CMB is consistent with the model. To recover something very roughly consistent with their result, we followed a different approach: we assumed that, since the Universe is almost FLRW up to the time of recombination, one can use the standard relation between $a$ and $z$, $a=1/(1+z)$, instead of Eq. (\ref{coordinate distance}).
The underlying idea is simple: the multipole $l_a$ should be affected only by how the early time physics, imprinted in the CMB, is projected to an observer today. The projection instead is affected by  backreaction, since the path of photons from the last scattering surface is perturbed by the growing inhomogeneities. This argument seems to be reasonable but it actually implies a theoretical inconsistency: together with $a_{D_0}= a_0 =1$, we are assuming a further boundary condition for $a$: the scale factor at recombination is fixed to $1/1+z^*$ both in FLRW and in backreaction.
 If none of the two boundary conditions is relaxed, this leads to a wrong estimate of the sound horizon and consequently of the multipole $l_a$. The latter is used for the computation of the positions of the peaks and is also one of the three shift parameters. The same problem affects also the shift parameter $R$, that also depends on the recombination scale factor, and obviously the third shift parameter, $z^*$. This would explain why both the CMB shift parameters and the peaks positions are affected by the same behavior.
  The problem is then understanding which one between $a^*$ and $z^*$ is the fundamental parameter for the recombination physics, from which the other one has to be determined. 
In our opinion the fundamental parameter should be the size of the Universe, $a^*$, rather than the redshift, because in the Boltzmann equations  it is the density $n a^3$ of the species present in the baryon-photon plasma that determines the time of recombination.
We tried hence to find an initial condition $a_{D_0}$ such that $a_D(z^*)=a^*$. This boundary-value problem translates into the integral equation
\begin{equation}
2\log(a_{D_0})=\int^{a_{D_0}}_0\frac{k'_D(x,a_{D_0})r^2(x,a_{D_0})}{1-k_D(x,a_{D_0}r^2(x,a_{D_0}))}dx
\end{equation}
where the unknown is $a_{D_0}$. Unfortunately, we were unable to solve this equation.
 Moreover, if $a^*$ is indeed the fundamental parameter,  we are not allowed to use Doran \& Lilley's fitting formulae, because they give the expression of $z^*$ in the standard FLRW cosmology, assuming implicitly $a_0=1$.  This reflects into a wrong prediction of the sound horizon and consequently in an error in predicting the fundamental multipole $l_a$.
Once this problem is solved, one should also deal with the issue that 
coarse-graining radiation-dominated Universe has extra effects, due to coarse-graining the radiation pressure inhomogeneities \cite{Buchert:2001sa}. The $a_D^{-4}$ dependence  of the radiation density in $H_D^2$ is built in analogy with standard cosmology and its effects are negligible on the likelihood contours, but at early times, when radiation is no more negligible, the effect of coarse graining may, in principle, be important.
Finally, the assumption that a perturbed FLRW metric holds up to recombination could also be questioned, because we cannot be sure whether the coarse graining effect is small enough to be completely negligible and to leave the CMB theory unaffected.
%
To conclude, given all the difficulties explained above,
our approach in this work is to simply assume that at the time of recombination the standard cosmological model is still a good approximation, and to check that our results are consistent with the other data sets. If an inconsistency appears, as will be the case, although it will be impossible in principle to tell whether this is due to the model being excluded by the data or by  a wrong analysis of the data, we will be inclined towards the second explanation.

Interesting proposals for using consistently CMB parameters in non-FLRW models are given in  \cite{Clarkson:2010ej} and  \cite{Vonlanthen:2010cd}, but unfortunately none of them helps us solve our problem\footnote{ 
\cite{Clarkson:2010ej} shows how to write some CMB parameters in terms of local quantities only, and in terms of temperatures rather than $a$ or $z$.  Temperatures should be more appropriate in inhomogeneous models because they can be derived through the local radiation density. However  while in the case of the void model treated in \cite{Clarkson:2010ej} it is possible to determine the exact relation between observables at the time of decoupling and observables today, in our case a correspondence between the bare scale factor $a$ and the averaged one $a_D$ is made so that  our metric looks like a FLRW model if seen as a function of $a_D$, but differs from it in the relation between (averaged) scale factor and (averaged) redshift. For this reason it would be arbitrary to decide how the measured temperatures are related to $a_D$ or $z_D$.
\cite{Vonlanthen:2010cd} finds model-independent parameters from the CMB, but this does not help us because even if we had perfectly model-independent variables at the time of decoupling, we could not correctly relate these to present-day parameters, because two incompatible boundary conditions are chosen.}.

\begin{figure}[!ht]
\begin{center}
\includegraphics[width=0.3\textwidth]{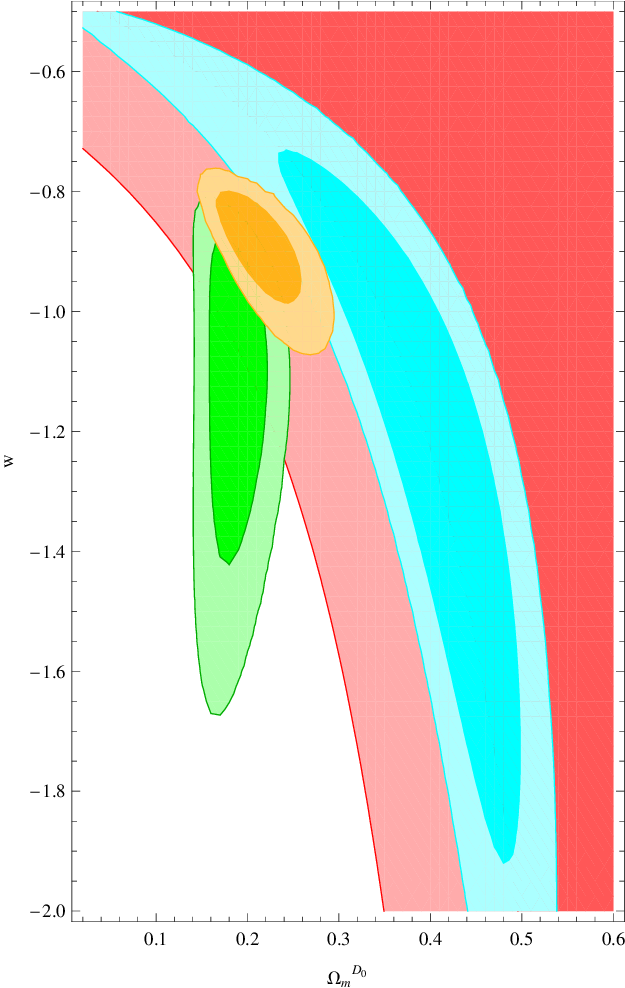}
\includegraphics[width=0.3\textwidth]{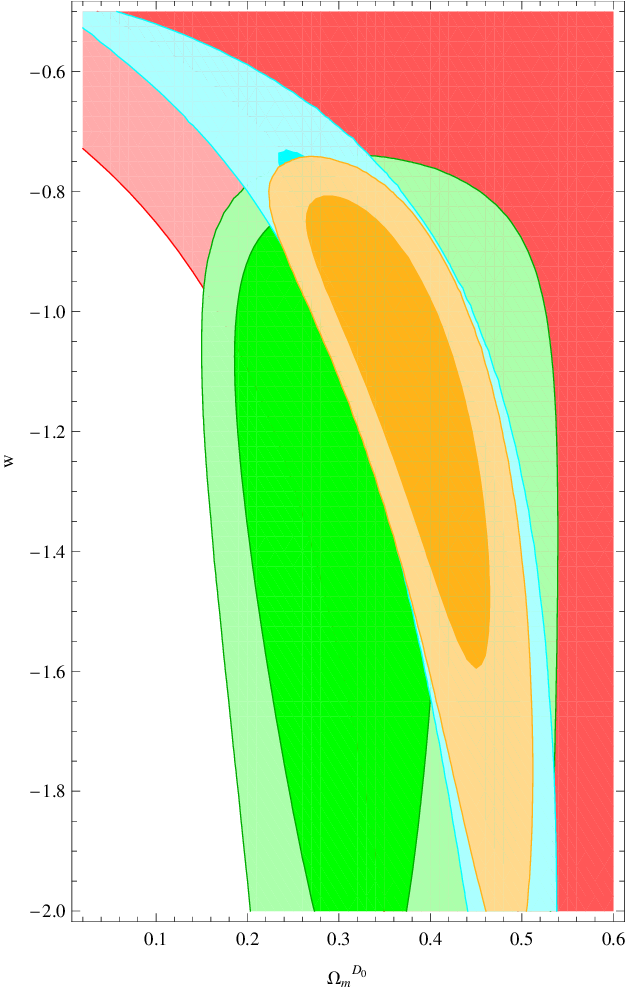}
\includegraphics[width=0.3\textwidth]{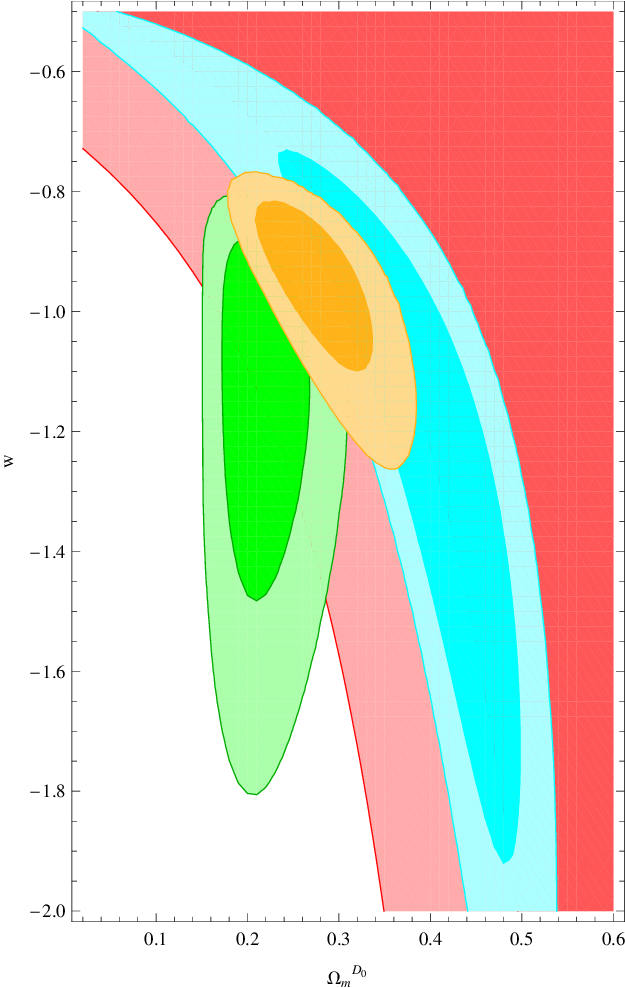}
\includegraphics[width=0.3\textwidth]{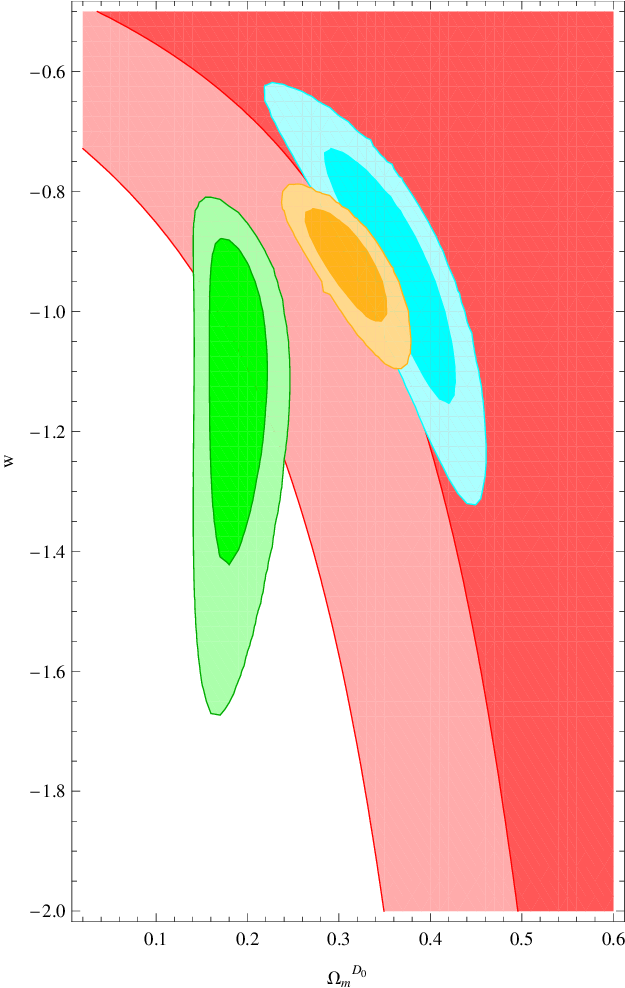}
\includegraphics[width=0.3\textwidth]{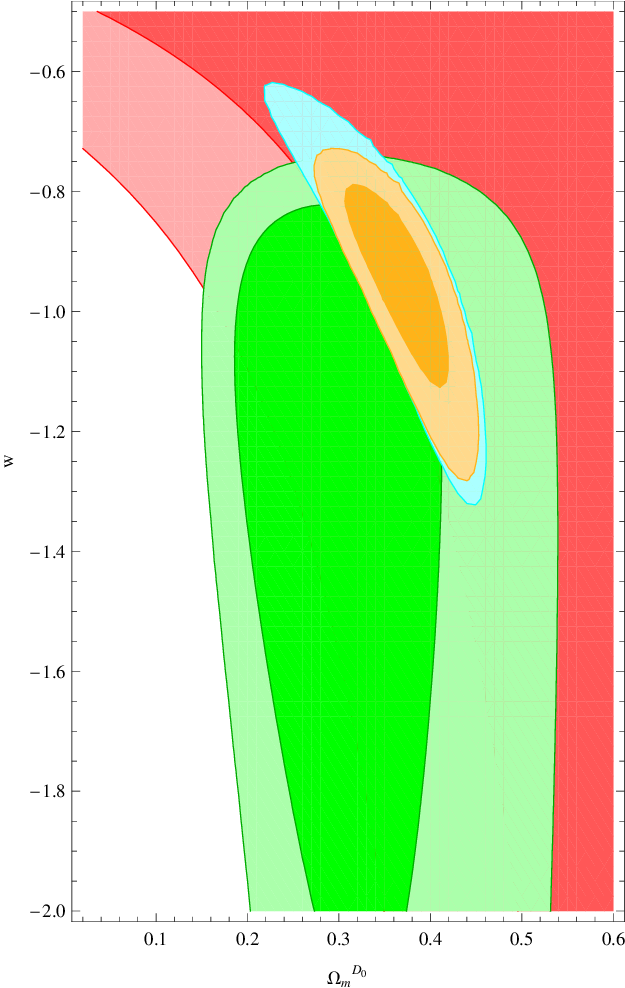}
\includegraphics[width=0.3\textwidth]{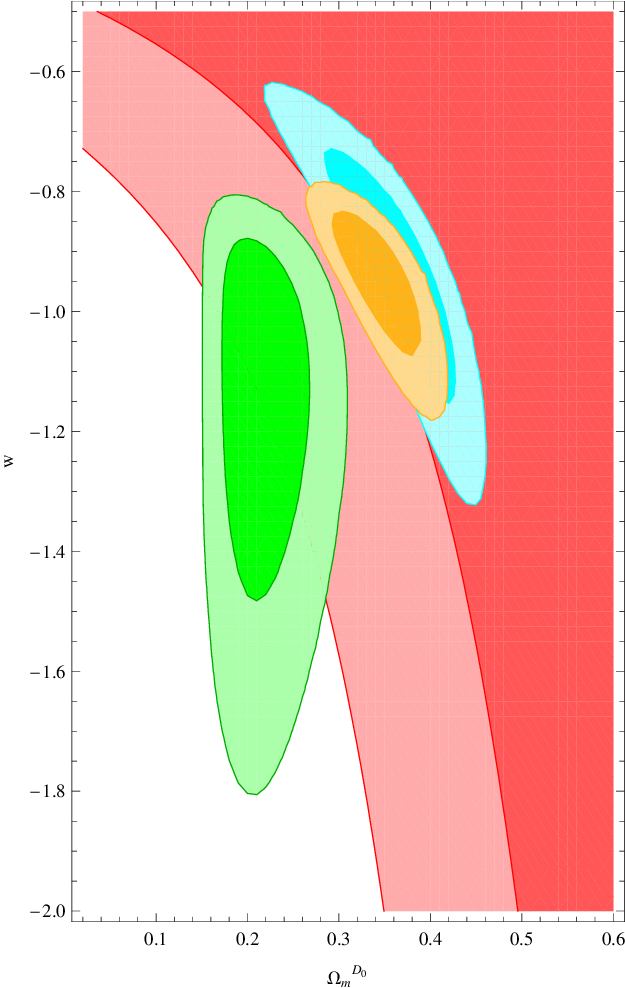}
\caption{$1$ and $2\sigma$ likelihood contours for the effective parameters $\Omega_m^{D_0}$ and $w$ for the backreaction model. 
Red fields correspond to clusters. Green fields comes from CMB data: \emph{WMAP-9} constraints from the CMB shift parameters are shown in the left panels, \emph{WMAP-9} constraints from the position of the peaks are shown in the central ones, constraints from \emph{Planck} are shown in the right panels. The blue fields correspond to SnIa: \emph{SNLS} constraints are shown in the upper panels, \textit{Union2.1} constraints are in the lower ones. The yellow contours are joint constraints.}
\label{fig:joint-contours}
\end{center}
\end{figure}

Combined constraints are shown in Fig.  \ref{fig:joint-contours}.
Here blue fields indicate SnIa constraints (upper and lower panels correspond to \emph{SNLS} and \textit{Union2.1} data, respectively), green fields indicate CMB constraints (left panels correspond to constraints from the shift parameters, while central and right panels correspond to constraints from the position of the peaks and dips of the CMB spectrum from  \emph{WMAP-9} and \emph{Planck} data, respectively), red fields indicate cluster constraints, and yellow fields correspond to joint constraints.

Since constraints from clusters and SnIa data move slightly in the direction of a  Universe with higher $\Omega_m^{D_0}$ while those from CMB data move in the opposite direction, the region of the parameter space where the likelihood functions overlap  becomes narrower for backreaction with respect to what happens for the standard FLRW case, and the different datasets become less compatible. The most interesting cases regards CMB shift parameters or \emph{Planck} data, which produce likelihoods that do not overlap at $2 \sigma$ with \textit{Union2.1} data likelihoods.
There are two possible interpretations of the latter. Either the template backreaction metric of Eq. (\ref{template}) is only marginally compatible with the data that we have used, or, as we believe, the inconsistency on the boundary conditions $a_{D_0}=a_0=1$ starts producing noticeable effects and cannot be neglected anymore.
In the former case, our work suggests that future data may be able to rule out the backreaction template metric, while in the latter case, more care will be needed in correctly defining observables in the backreaction context, since constraints may be strongly affected by them. 
In particular, as explained above, it is critical to understand well the meaning and use of $z^*$, which enters directly the calculation of the multipole $l_a$, on which CMB constraints are based.
It is interesting to understand whether the clusters dataset used here for the first time to constrain a backreaction cosmology, brings relevant improvements. 
\begin{figure}[!ht]
\begin{center}
\includegraphics[width=0.3\textwidth]{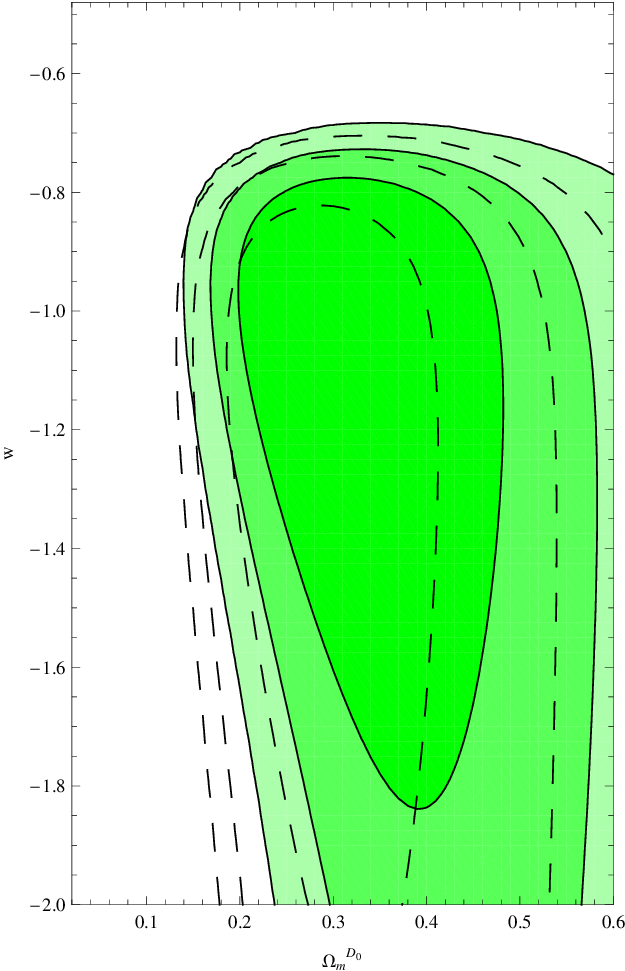}
\includegraphics[width=0.3\textwidth]{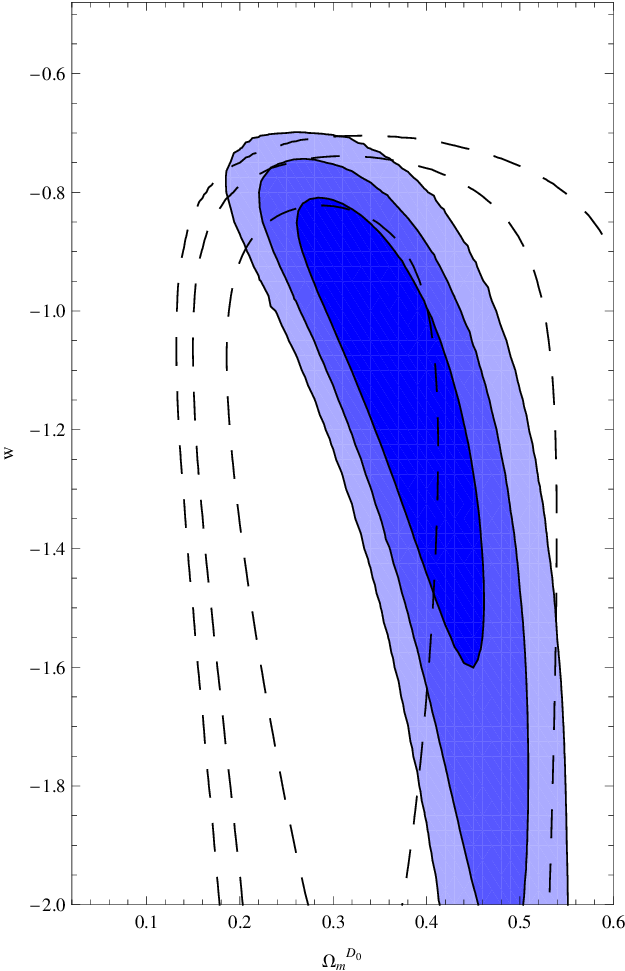}
\includegraphics[width=0.3\textwidth]{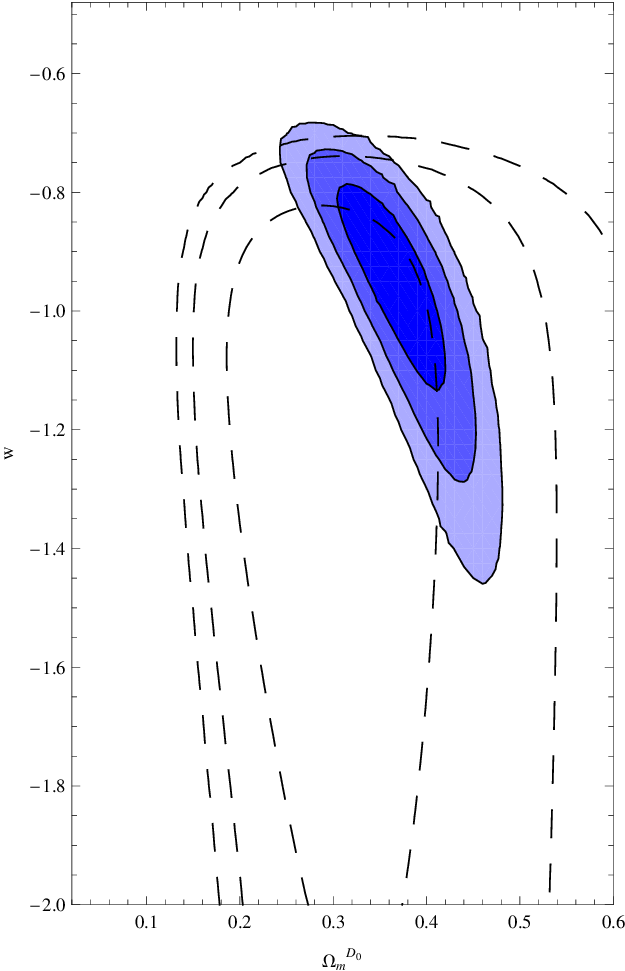}
\includegraphics[width=0.3\textwidth]{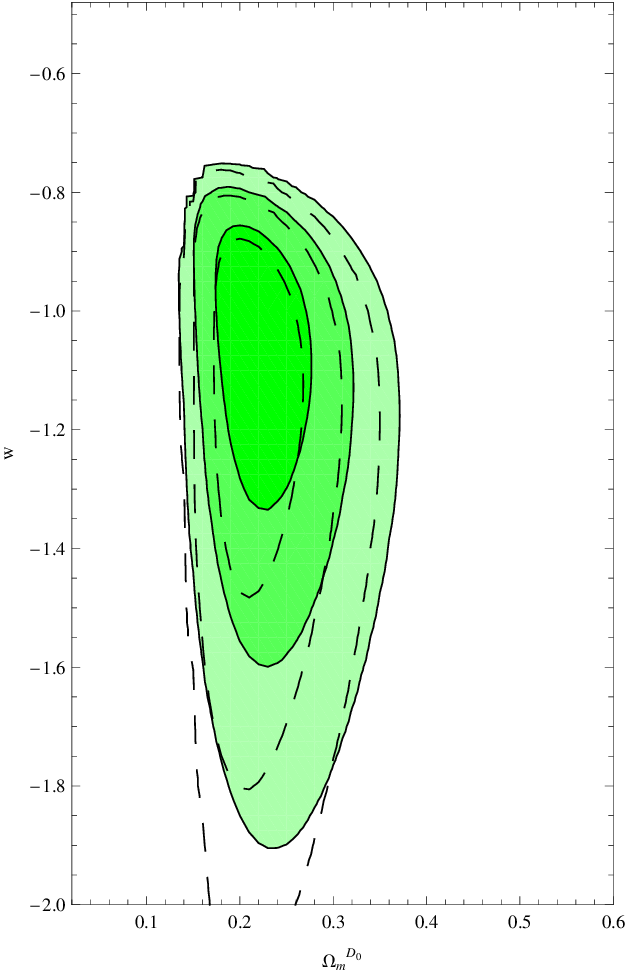}
\includegraphics[width=0.3\textwidth]{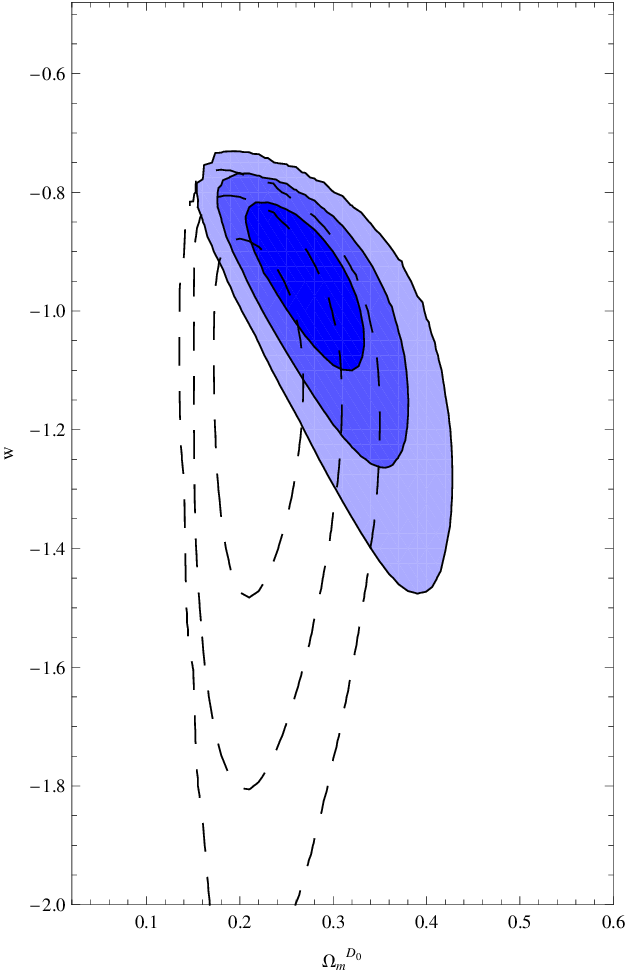}
\includegraphics[width=0.3\textwidth]{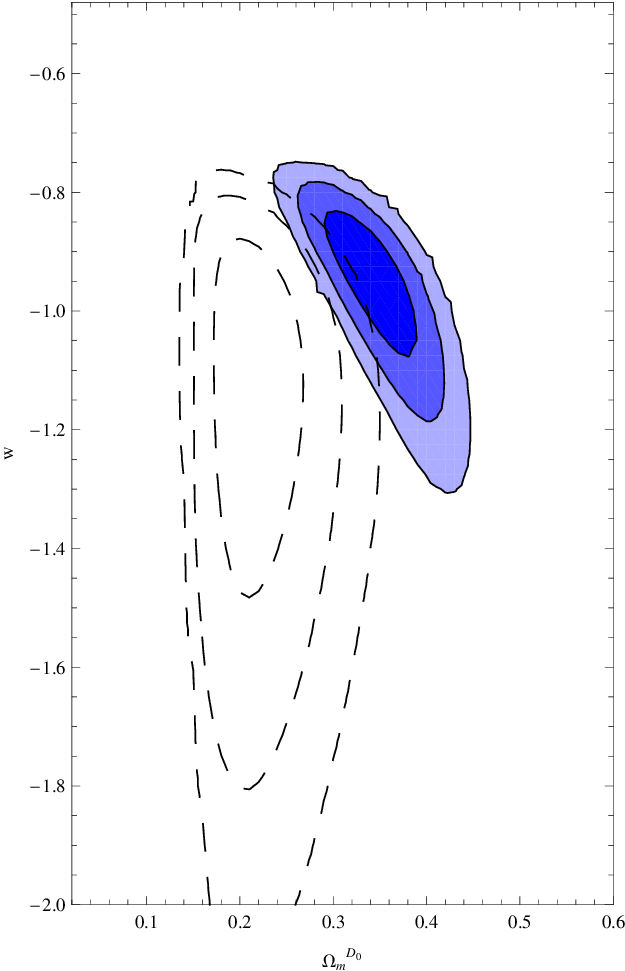}
\caption{Comparison between constraints from the CMB peaks alone and combined ones, for the backreacted Universe. Blank contours refers to CMB data alone. Filled contours are joint constraints. Contours in the upper panels comes from \emph{WMAP-9} data, those in the lower ones comes from \emph{Planck}. Left panels: combination with cluster data. Central panels: combination with \emph{SNLS} data. Right panels: combination with \textit{Union2.1} data.}
\label{fig:cmb}
\end{center}
\end{figure}
Hence, in Fig.~\ref{fig:cmb} we show how CMB peaks constraints improve when combined with constraints derived from clusters. For comparison, also the improvement on CMB constraints from SnIa is shown.
The upper panels show \emph{WMAP-9} data, while lower panels show \emph{Planck} data. 
As can be seen in the left panels, the improvement obtained by adding clusters data is not very strong. On the contrary, improvements induced by adding SnIa data (shown in the central and right panels, corresponding to \emph{SNLS} and \emph{Union2.1} data, respectively) are much more significant. In particular, \emph{SNLS} data are less precise and combined constraints are looser, while the \textit{Union2.1} data provides tighter constraints. 
As regards \emph{WMAP-9} data, we note that all combined constraints overlap constraints from CMB data alone, indicating that backreaction is compatible with these data sets. This is also due to the large errors associated to the position of the peaks caused by our fitting procedure.
Regarding \emph{Planck} data, they are still compatible with  \emph{SNLS} data, but when combining them with \emph{Union2.1} data,
the $1\sigma$ regions of the joint constraints do not overlap those corresponding to CMB data alone, and, as explained above, this may either imply that the model is not compatible with these observations or rather that observations are precise enough to necessitate a refinement of the approximations made until now.

\begin{figure}[!ht]
\begin{center}
\includegraphics[width=0.3\textwidth]{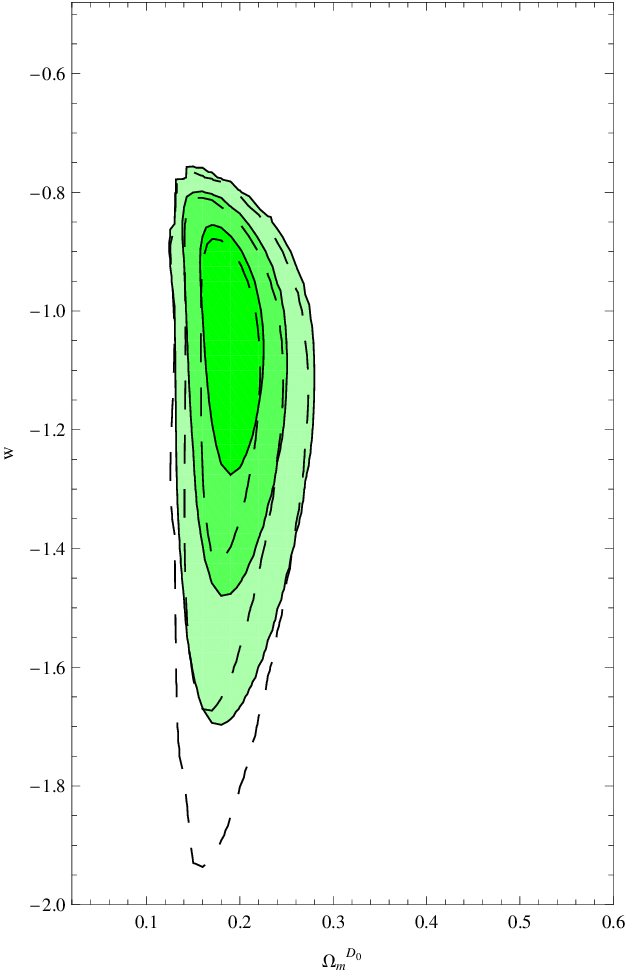}
\includegraphics[width=0.3\textwidth]{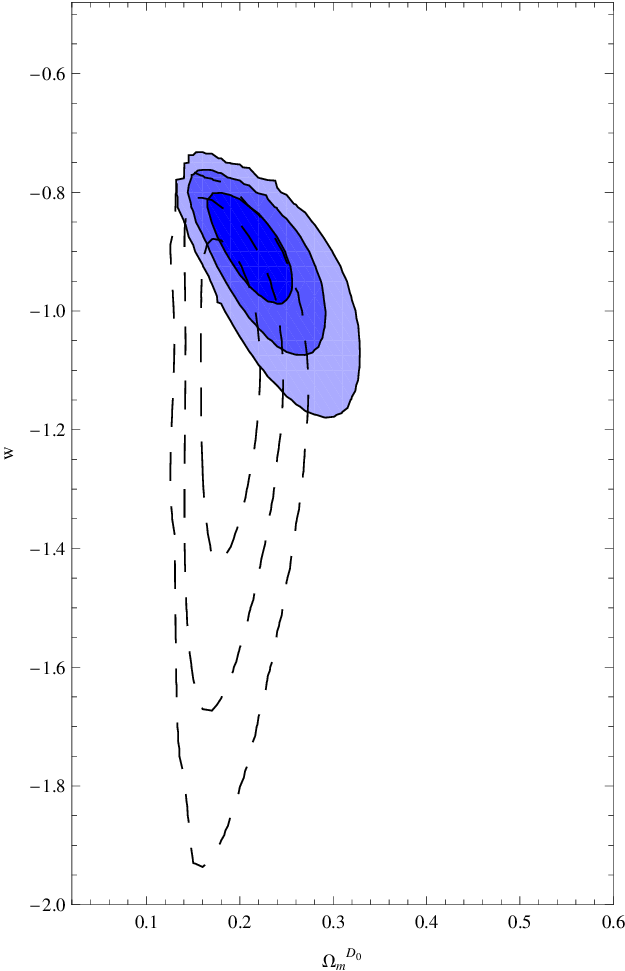}
\includegraphics[width=0.3\textwidth]{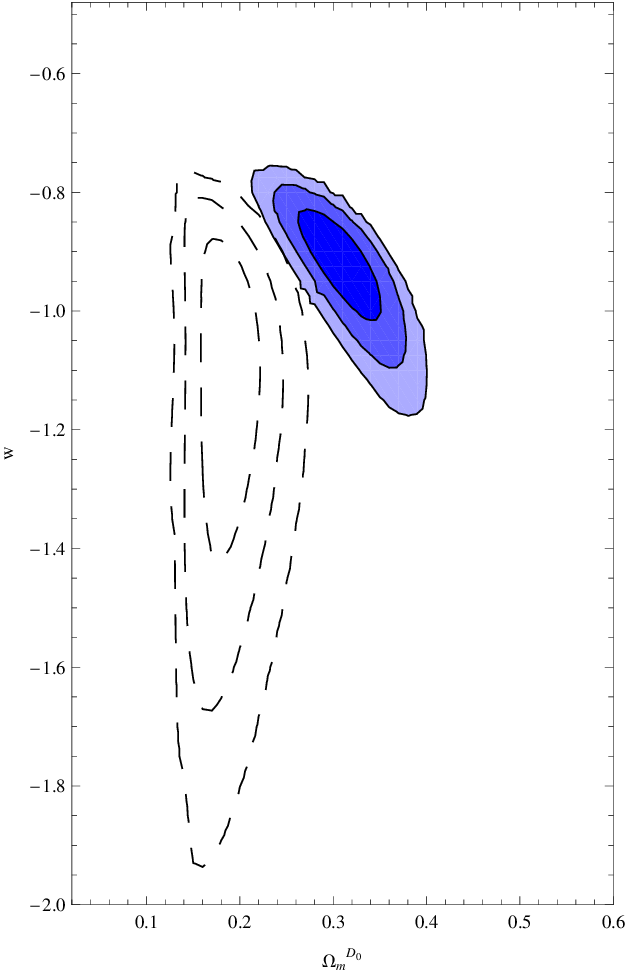}
\caption{Comparison between constraints from the CMB shift parameters and combined ones for the backreacted Universe. Blank contours refers to CMB data alone. Filled contours are joint constraints. Left panel: combination with cluster data. Central panel: combination with \emph{SNLS} data. Right panel: combination with \textit{Union2.1} data.}
\label{fig:cmbshift}
\end{center}
\end{figure}

\begin{figure}[!ht]
\begin{center}
\includegraphics[width=0.5\textwidth]{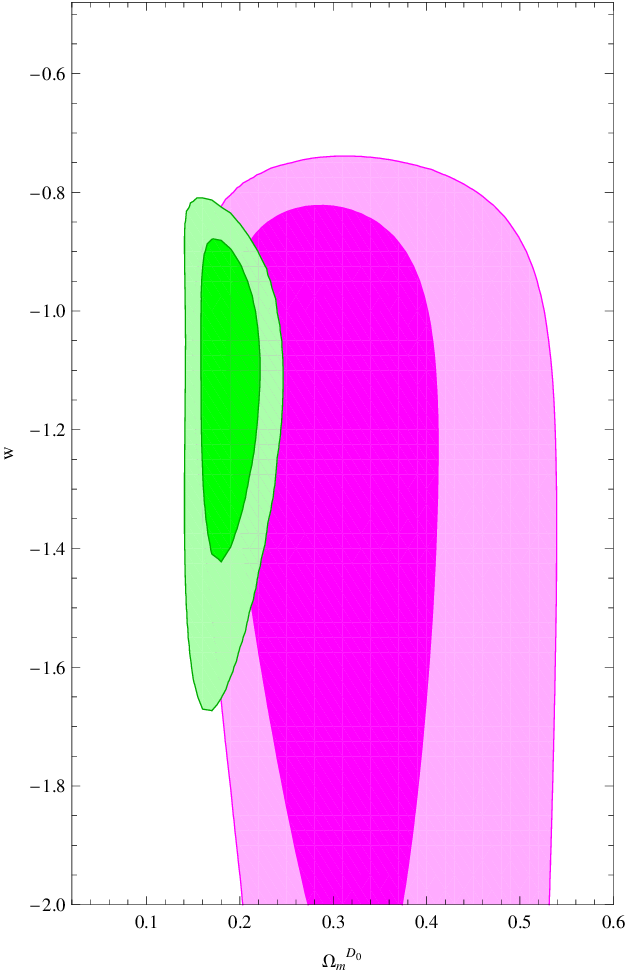}
\caption{Comparison between constraints provided by the position of the CMB peaks and dips (purple fields) and those given by the CMB shift parameters (green fields) for the averaged model. Only \emph{WMAP-9} data were used.}
\label{fig:cmb-backreaction}
\end{center}
\end{figure}

In Fig.~\ref{fig:cmbshift} we show the same as in the bottom panels of Fig.~\ref{fig:cmb}, but for CMB shift parameters from \emph{WMAP-9}. The latter constraints are tighter than those from \emph{Planck}'s peak positions. The right panel shows clearly that also in this case the \textit{Union2.1} data provide constraints only marginally compatible with CMB shift data.
Finally, from the right panel of Fig.~\ref{fig:clusters-improvement} we had noted from \emph{WMAP} data that  there is concordance between the CMB peaks and CMB shift parameters methods in the standard FLRW framework, since the two sets of constraints overlap very well. This  is natural, given that the likelihoods are not independent since they are based on the same data set and in particular since the multipole $l_a$ is used in both procedures. 
Surprisingly instead, when considering backreaction, the two methods give results that are not concordant, as shown in Fig.~\ref{fig:cmb-backreaction}. The $1\sigma$ regions relative to the CMB shift parameters do not overlap completely with the corresponding region from the CMB peaks. We claim that this is the mark of the inconsistency we have described earlier and that the main  problem consists in correctly predicting $z^*$. Indeed if only the theoretical computation of $l_a$ were affected by an inconsistency both constraints should approximately modify likewise. The strong modification which affects only contours given by the shift parameters can be explained by stating that also the theoretical predictions of $z^*$ and $R$ are corrupted. Since $R$ implicitly depends on $z^*$ too, we are led to guess that the evaluation of the decoupling epoch in the averaged model requires more care and better approximations.

\section{Conclusions}\label{sec:conclusions}

We tested the backreaction model based on the template metric  of Eq. (\ref{template}) suggested in \cite{Larena:2008be} against recent data sets. The morphon field creates a correspondence between an averaged Universe based on Buchert's averaging procedure on spatial slices and a standard flat FLRW spacetime filled with DE  of constant equation of state. We may distinguish among the two models by assuming a template metric for the averaged Universe and testing it against present data. 

In \cite{Larena:2008be} the authors published constraints on the effective cosmological parameter $\Omega_m^{D_0}$ and the scaling index of the backreaction $n$ by combining  SnIa of the \emph{SNLS} the position of the first two peaks and the first dip of the CMB power spectrum obtained using \emph{WMAP-3} data.
We updated their analysis by using the SnIa of the \textit{Union2.1} catalog and by employing for the first time in the context of backreaction the set of angular diameter distances of galaxy clusters from Bonamente \cite{Bonamente:2005ct}. We found that while the \textit{Union2.1} data improve noticeably the constraints from the \emph{SNLS} due to the higher number of SnIa, no useful information is added by clusters, since the errors are heavily affected by uncertainties on modeling the geometry of the clusters. Nevertheless we found interestingly that clusters data confirm the behavior followed by the constraints shown in Fig. 2 of \cite{Larena:2008be}, suggesting that future surveys of galaxy clusters, and refined modeling errors, may be useful for constraining the cosmological parameters and may help in distinguishing between an effective solution of the Einstein field equation and the standard homogeneous and isotropic FLRW.

We also updated the likelihood analysis based on data from the CMB. We  used two different sets of observables. First we followed the path suggested in \cite{Larena:2008be}, and used the positions of the first three peaks and the first dip of the CMB temperature power spectrum. We found that adding a point corresponding to the third peak to \emph{WMAP-3} data improves the constraints despite its big error, which was derived by a not very refined fitting procedure. We  applied the same fitting procedure to \emph{WMAP-9} and \emph{Planck} data, extracting two sets of data corresponding to the position of the first three peaks and the first dip of the CMB spectrum. 
As an alternative to this observable, we used the CMB shift parameters provided by \emph{WMAP-9}. We found that the latter dataset gives better constraints than the former. This is partly due to the poor estimate of errors of the former dataset.
\emph{Planck} data would be sufficiently precise to give good results, but the fitting procedure we used for extracting the positions of the peaks produces large errors. In the future we plan to improve our estimate by using additional parameters related to the separation points between peaks and by using a refined Monte Carlo Markov Chain technique.
The analysis of the CMB provides good constraints for the FLRW spacetime. However, when considering backreaction, we find a behavior opposite to what shown in \cite{Larena:2008be}. As a consequence, the \emph{WMAP-9} CMB shift likelihood contours  only marginally overlap those produced by  \emph{SNLS} data and the overlap reduces further when \textit{Union2.1} data are used. 
Following \cite{Rosenthal:2008ic}, we tried to explain this behavior in terms of inconsistencies in the computation of the redshift of recombination, caused by the non trivial relation between the effective scale factor and the effective redshift before and after the recombination epoch. 
In particular, we believe that the inconsistency arises due to the assumption of standard boundary conditions ($a_0=a_{D_0}=1$ together with $a^* = a^*_D$) for both the FLRW spacetime and the backreacted model. 
We tried to demonstrate that the approximation suggested in \cite{Larena:2008be} of a Friedmannian evolution for any domain $D$ in the backreacted Universe up to the recombination epoch is not well posed. 
This led us to suggest that the correspondence between an averaged model and a standard flat FLRW filled with DE should be formulated for suitable boundary conditions today. 
We noted that, in the standard cosmology, fixing the recombination epoch means fixing the size of the Universe, and we suggested that the main role is played by the latter, since it has a thermodynamical meaning related to the  number density of species.  Unfortunately, our attempts of solving the boundary conditions problem in terms of $a$ have failed, as we ran into an integral which we were unable to solve. Finally we argued that the inconsistency should affect the likelihood analysis because the effective scale factor $a_D(z_D)$ cannot correspond to $a(z) = (1+z^*)^{-1}$ at the recombination epoch $z^*$.

SnIa and clusters provide compatible constraints and a higher best fit value of $\Omega_m^{D_0}$ is found if backreaction is considered. 
Current CMB data, on the other hand, clearly tell us that the inconsistency related to how we treat the CMB physics in order to avoid the complication of rewriting all the microphysics of the photon-baryon fluid on a spacetime described by the template metric cannot be neglected anymore. Mistakes induced by this approach are now of the same order of the experimental errors, and predictions based on current data are strongly affected by them.

\appendix

\section{Fitting procedure}\label{app:peaks}

Here we discuss the fitting procedure we used for finding the position of the peaks and dips of the CMB. We applied the same scheme to \emph{WMAP-9} raw data and to the combined power spectrum measured by \emph{Planck}.

The \emph{WMAP-9} data consist of a list of $l(l+1)C_l/(2\pi)$ and the corresponding statistical error as a function of the multipole $l$,  from $l=2$ up to $l=1200$.
We arbitrarily divide the range $l\in[2,1200]$ into subsets corresponding to a rough estimation of the extension of each peak or dip: 
\begin{itemize}
\item First peak: $l\in[50,350]$
\item First dip: $l\in[300,550]$
\item Second peak: $l\in[400,700]$
\item Third peak:$l\in[650,950]$.
\end{itemize}
We decided to take overlapping intervals for the different peaks, and we checked that the different fitting parabolas always cross each other in the overlapping regions of neighboring ranges.
After that, we model each peak or dip with a 3-parameter curve. The first peak is modeled with a Gaussian, while the other two peaks and the first dip are modeled with parabolas: $p(x)=ax^2+bx+c$. For the first peak, we fit the logarithm of the spectrum, in order to transform the Gaussian fit into a standard parabolic fit.
Once the best fitting values of $a$, $b$ and $c$ are found for each peak or dip, we estimate its position assuming that it is well approximated by the position of the vertex $v=-b/2a$.

In order to achieve a statistical sample of the position of the vertex, once we have estimated the position of the vertex, we randomly add a different random number to each data point and we refit the data. The range in $l$ is held fixed and the random numbers are extracted from a Gaussian distribution with null mean and standard deviation corresponding to the statistical error associated to each data point. 
The refitting procedure is done at least $10^5$ times. This number of iterations  ensures that the vertices follow a Gaussian distribution, hence that the main uncertainty in the position of the peak is truly statistical.
The final position of the $n$-th peak (or dip) is computed by taking the mean of the sample, and the corresponding error is estimated as the sample standard deviation.

We tried to estimate the error introduced by fixing the fitting range in the following way.  We let the boundaries of the fitting range fluctuate randomly. We assume that each boundary point can fluctuate with a Gaussian distribution with null mean and $\sigma$ corresponding to the 10\% of the width of the interval. Again we note that this choice is arbitrary, but it should not affect too much our estimate. 
We refit the data each time the boundaries of the fitting range are changed. We then estimate the error on the position of each peak as the standard deviation of the corresponding sample, which contains again $10^5$ realizations.
We find that the uncertainty on the position of each peak introduced by fixing the fitting range is significant only for the first peak and the first dip because it is of the same order of magnitude of the statistical uncertainty predicted by the fitting procedure holding the extrema fixed. Hence we decided to add to the final error on the first peak and on the first dip this contribution  in quadrature, assuming that the errors are uncorrelated.

\begin{figure}[!ht]
\begin{center}
\includegraphics[width=0.4\textwidth]{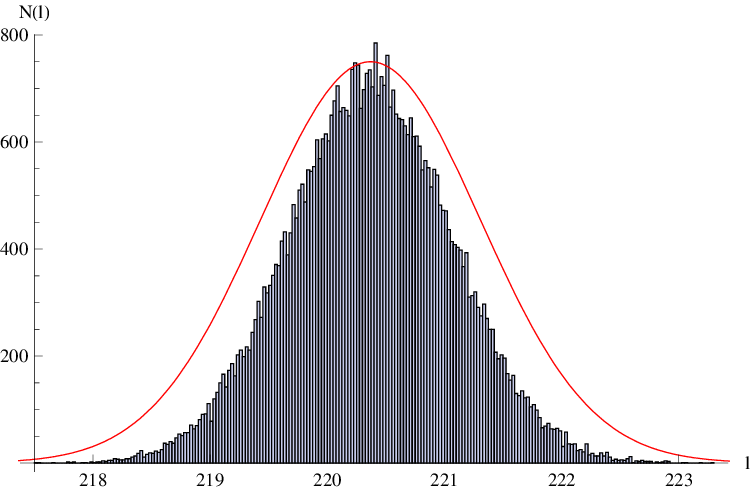}
\includegraphics[width=0.4\textwidth]{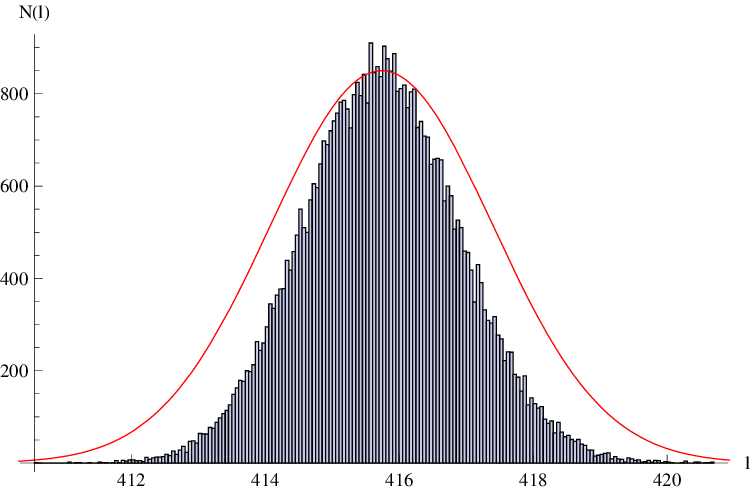}
\includegraphics[width=0.4\textwidth]{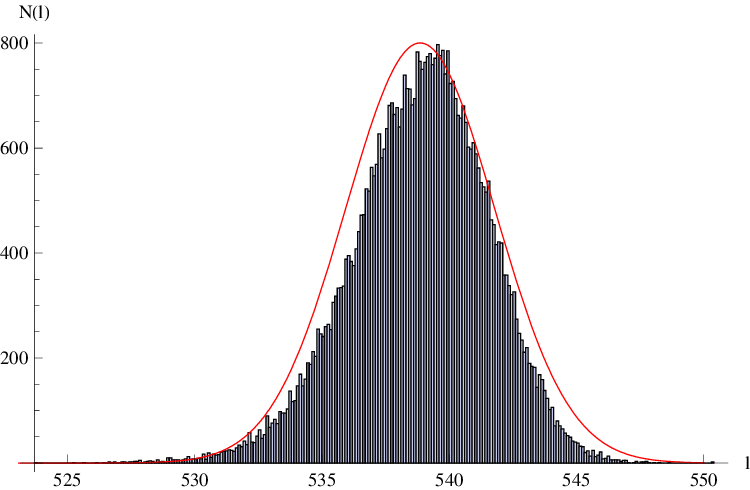}
\includegraphics[width=0.4\textwidth]{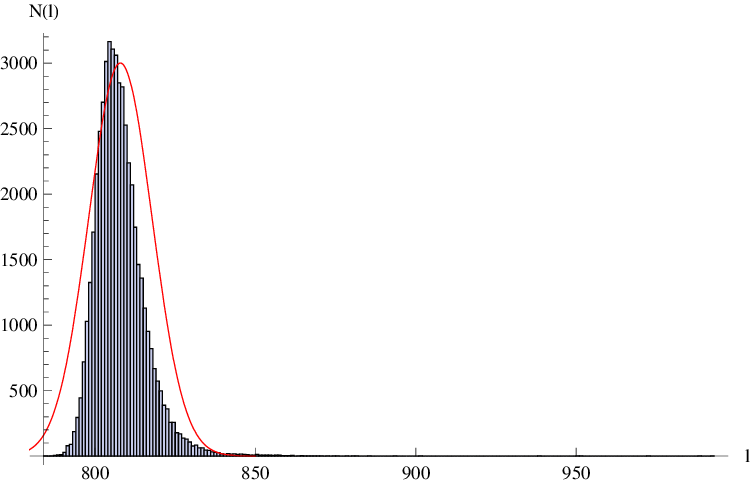}
\caption{Fitting procedure of \emph{WMAP-9} data: distribution of the estimator of the position of the peaks and dips. Upper left: first peak. Upper right: first dip. Lower left: second peak. Lower right: third peak}
\label{fig:peaks_wmap}
\end{center}
\end{figure}

In Fig.~\ref{fig:peaks_wmap} we show the histograms of the values of the vertices found from fitting the three peaks and the first dip. For each histogram the corresponding Gaussian is superimposed for an easy visualization. The Gaussians corresponding to the first peak and to the first dip are slightly wider than the envelope of the histogram, since their variance contains the contribution coming from fixing the fitting ranges.

\begin{figure}[!ht]
\begin{center}
\includegraphics[width=0.4\textwidth]{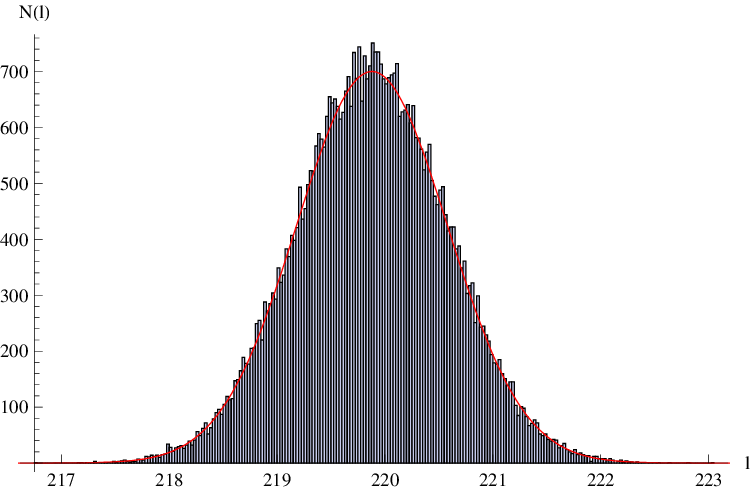}
\includegraphics[width=0.4\textwidth]{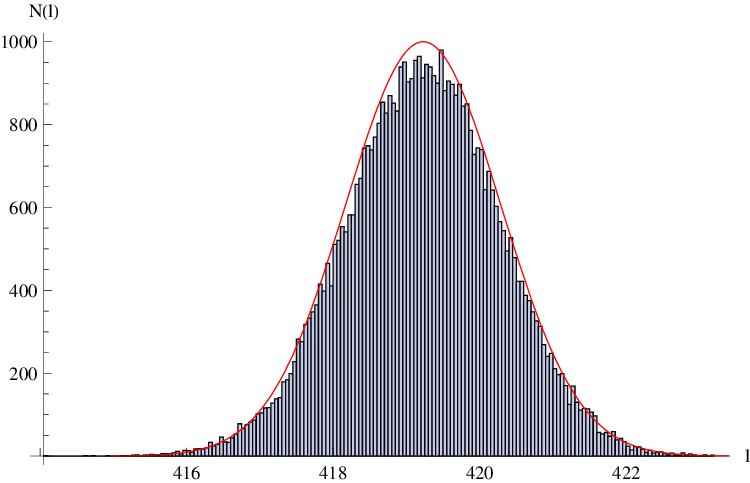}
\includegraphics[width=0.4\textwidth]{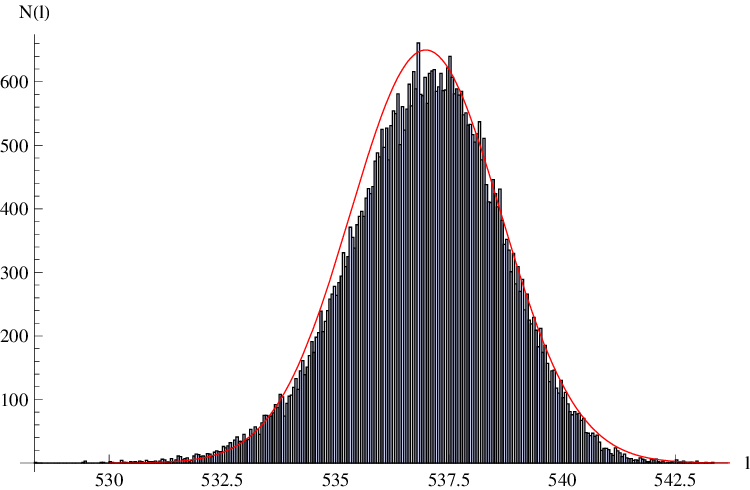}
\includegraphics[width=0.4\textwidth]{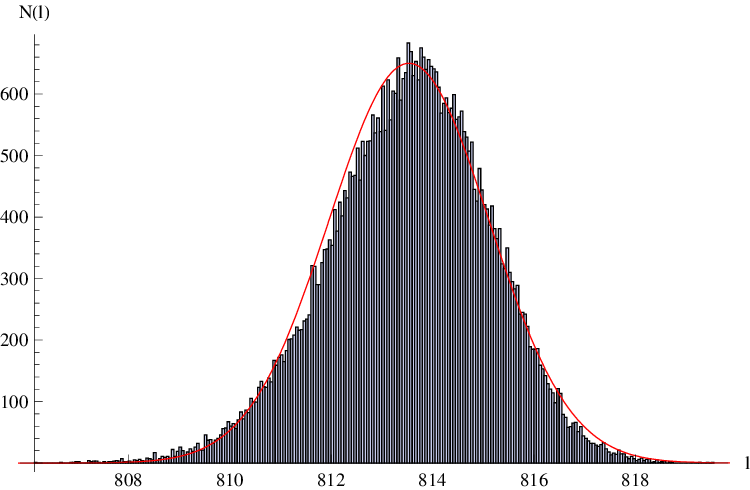}
\caption{Fitting procedure of \emph{Planck} data: distribution of the estimator of the position of the peaks and dips. Upper left: first peak. Upper right: first dip. Lower left: second peak. Lower right: third peak.}
\label{fig:peaks-planck}
\end{center}
\end{figure}

We applied the same procedure to the \emph{Planck} data. Here we note that the statistical uncertainty of each data point is drastically reduced due to the binning procedure employed by \emph{Planck}. We assumed the same ranges we used for fitting \emph{WMAP-9} data, but no error is computed relative to fixing the range, since the data points  are too distant in $l$.
In Fig.~\ref{fig:peaks-planck} we show the histograms corresponding to the predicted values of the position of each vertex of the fitting parabolas and the corresponding Gaussian.

A comparison between Fig.~\ref{fig:peaks_wmap} and Fig.~\ref{fig:peaks-planck} shows that the position of the third peak is estimated with better precision if \emph{Planck} data are used. The predicted positions of the third peak derived from \emph{WMAP-9} data distribute quite asymmetrically, while the position derived from \emph{Planck} data shows a more symmetric distribution. This is due to the fact that uncertainties on \emph{WMAP-9} raw data increase drastically for $l>700$ and the prediction of the position of the third peak may be biased towards lower values of $l$, corresponding to lower errors.

\begin{figure}[!ht]
\begin{center}
\includegraphics[width=0.7\textwidth]{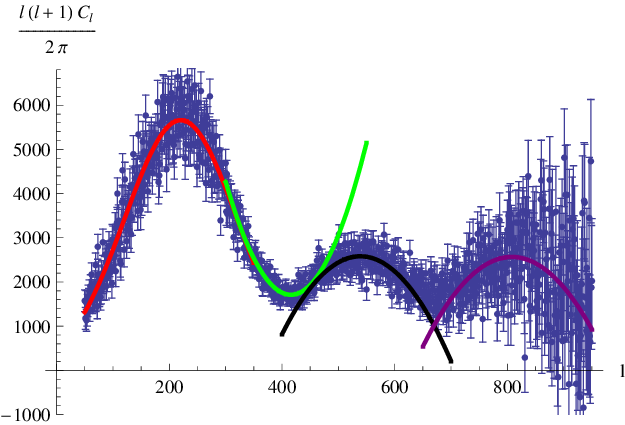}
\includegraphics[width=0.7\textwidth]{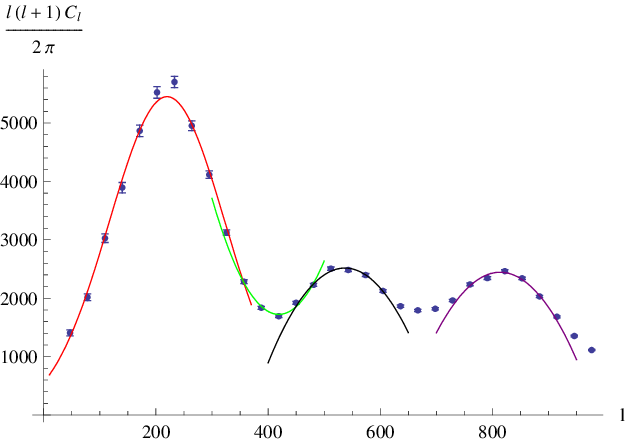}
\caption{Fitting procedure of CMB data: superposition of raw data and fitting curves. Upper panel: \emph{WMAP-9} data. Lower panel: \emph{Planck} data}
\label{fig:spectra-fits}
\end{center}
\end{figure}

Finally,  in Fig.~\ref{fig:spectra-fits} we show the fitting parabolas superimposed on the measured spectra.

\acknowledgments
We acknowledge Luigi Guzzo, Martin Kunz, Pierstefano Corasaniti, Thomas Buchert, Marco Bersanelli, Davide Bianchi and Alida Marchetti for useful discussions.
E.~M. was supported by the Spanish MICINNs Juan de la Cierva programme (JCI-2010-08112), by CICYT through the project FPA-2012-31880, by the Madrid Regional Government (CAM) through the project HEPHACOS S2009/ESP-1473 under grant P-ESP-00346 and by the European Union FP7 ITN INVISIBLES (Marie Curie Actions, PITN- GA-2011- 289442). E.~M. also acknowledges the support of the Spanish MINECO's ``Centro de Excelencia Severo Ochoa" Programme under Grant No. SEV-2012-0249.

{}

\end{document}